\documentclass[preprint,review,3p,12pt,numbers]{elsarticle}

\usepackage{graphicx}
\usepackage{float}

\usepackage{amssymb}
\usepackage{amsmath}

\usepackage{lineno}
\usepackage{multirow}
\usepackage[table,xcdraw]{xcolor}
\usepackage{longtable,tabularx}
\usepackage{nomencl}
\makenomenclature
\usepackage{etoolbox}
\renewcommand\nomgroup[1]{%
  \item[\bfseries
  \ifstrequal{#1}{A}{Sets}{%
  \ifstrequal{#1}{B}{Parameters}{%
  \ifstrequal{#1}{C}{Binary variables}{%
  \ifstrequal{#1}{D}{0-1 continuous variables}{%
  \ifstrequal{#1}{E}{Continuous variables}{%
  \ifstrequal{#1}{F}{Abbreviations}{}}}}}}%
]}
\usepackage{color}
\newcommand{\blue}[1]{\textcolor{blue}{#1}}
\usepackage[colorlinks,linkcolor=black,anchorcolor=black,citecolor=black,urlcolor=blue]{hyperref}
\usepackage{array}
\newcolumntype{L}[1]{>{\raggedright\arraybackslash}p{#1}}
\usepackage{float}

\journal{}

\begin{document}

\begin{frontmatter}



\title{A production planning benchmark for real-world refinery-petrochemical complexes}


\author[inst1,inst2]{Wenli Du \corref{cor1}}
\ead{wldu@ecust.edu.cn}
\author[inst1,inst2]{Chuan Wang}

\affiliation[inst1]{organization={State Key Laboratory of Industrial Control Technology},
            addressline={East China University of Science and Technology}, 
            city={Shanghai},
            postcode={200237}, 
            country={China}}
\affiliation[inst2]{organization={Key Laboratory of Smart Manufacturing in Energy Chemical Process, Ministry of Education},
            addressline={East China University of Science and Technology}, 
            city={Shanghai},
            postcode={200237}, 
            country={China}}

\author[inst1,inst2]{Chen Fan}
\author[inst1,inst2]{Zhi Li}
\author[inst1,inst2]{Yeke Zhong}
\author[inst1,inst2]{Tianao Kang}
\author[inst1,inst2]{Ziting Liang}
\author[inst1,inst2]{Minglei Yang}
\author[inst1,inst2]{Feng Qian}
\author[inst1,inst2]{Xin Dai \corref{cor1}}
\ead{xindai@ecust.edu.cn}

\cortext[cor1]{Corresponding authors}


\begin{abstract}
To achieve digital intelligence transformation and carbon neutrality, effective production planning is crucial for integrated refinery-petrochemical complexes. Modern refinery planning relies on advanced optimization techniques, whose development requires reproducible benchmark problems. However, existing benchmarks lack practical context or impose oversimplified assumptions, limiting their applicability to enterprise-wide optimization. To bridge the substantial gap between theoretical research and industrial applications, this paper introduces the first open-source, demand-driven benchmark for industrial-scale refinery-petrochemical complexes with transparent model formulations and comprehensive input parameters. The benchmark incorporates a novel port-stream hybrid superstructure for modular modeling and broad generalizability. Key secondary processing units are represented using the delta-base approach grounded in historical data. Three real-world cases have been constructed to encompass distinct scenario characteristics, respectively addressing (1) a stand-alone refinery without integer variables, (2) chemical site integration with inventory-related integer variables, and (3) multi-period planning. All model parameters are fully accessible. Additionally, this paper provides an analysis of computational performance, ablation experiments on delta-base modeling, and application scenarios for the proposed benchmark.
\end{abstract}




\end{frontmatter}



\section{Introduction}
\label{sec:intro}
In response to intensifying market competition and the growing demand for energy conservation and emissions reduction, capital-intensive process industries, such as the petrochemical sector, must transition from traditional extensive production systems to more efficient and sustainable practices by restructuring production factors, enhancing resource utilization, and achieving smart and optimal manufacturing \citep{qian2017fundamental}. Within the petrochemical industry, refinery planning plays a critical role in improving enterprise-level efficiency across the entire production chain, directly impacting operational performance, profitability, and sustainability. As such, it constitutes a fundamental pillar of the smart manufacturing transformation. Its importance is further amplified by the industry's growing focus on integrating upstream refineries with downstream petrochemical plants, a strategy aimed at unlocking new business opportunities and enhancing overall process synergies \citep{KETABCHI201985}. Effective planning is vital for the efficient allocation of resources and the achievement of production goals without disruptions or inefficiencies. It requires the development of optimal production strategies that consider key factors such as raw material availability, market demand, production capacities, and various operational constraints \citep{li_development_2020}.

The investigation of production planning optimization problems can be traced back to the advent of linear programming (LP), which gained prominence following the introduction of the simplex method \citep{bodington_history_1990}. Since then, a plethora of studies have proliferated. Two primary difficulties are addressed across the literature from the industrial practices to theoretical research. One of them lies in the precise representation of resource flow and material transformation processes. It often requires balancing model accuracy with computational complexity. This trade-off is crucial, as highly detailed models can become computationally prohibitive. Another challenge stems from the intrinsic complexity of solving mixed-integer nonlinear programming (MINLP) problems, which are common in refinery planning due to the presence of discrete operating modes and pooling requirements. The expanding scale of integrated refinery-petrochemical planning intensifies the combinatorial explosion of potential solutions, making it harder to find optimal or near-optimal solutions efficiently.

\begin{figure}[htb]
    \centering
    \includegraphics[scale=0.7]{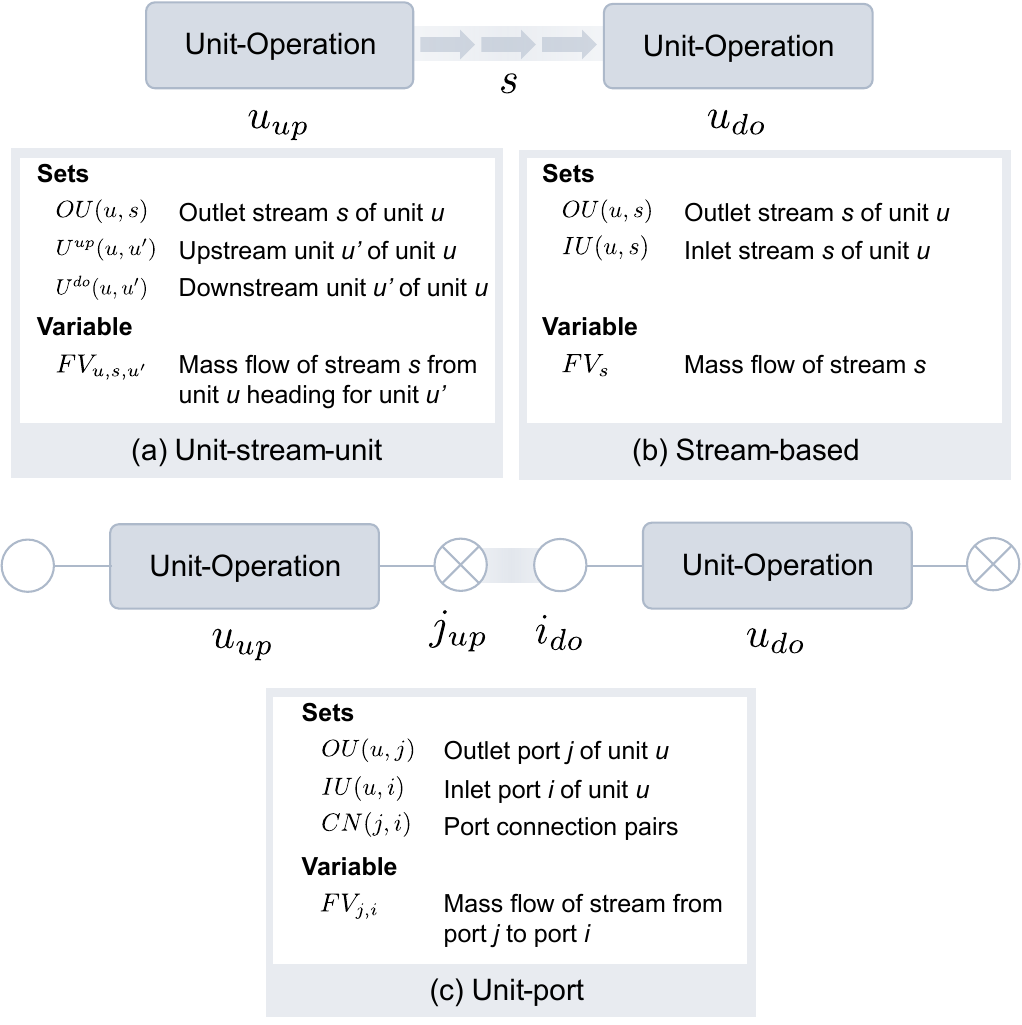}
    \caption{Key sets and variables to define a process network using different superstructures}
    \label{fig:supstrct}
\end{figure}

A precise and effective mathematical formulation is fundamental to achieving an optimal production plan. From a process-wide perspective, a unified and clear framework is required to accurately represent the refinery's topology, including well-defined identification of materials and flow directions. A common approach is to connect different units via explicit material streams \citep{NEIRO2004871,GAO200867,LI2021107361}. To establish a uniquely determined process network, either the upstream-downstream relationships between units or the inflow-outflow correspondences between units and streams must be specified. While this approach is intuitive, it may lack flexibility. An alternative method builds a topological network by linking two ports with inport or outport attributes. A notable example is the unit-operation-port-state superstructure (UOPSS) introduced by Zyngier and Kelly \cite{zyngier2012uopss}, in which each material flow is uniquely characterized by the inlet and outlet port pairs of distinct unit-operations. This framework allows for a flexible number of ports, enabling units to handle multiple tasks, which is an advantage not possible in models based on physical material streams. Studies indicate that this paradigm performs well in scaling up to model large-scale refinery scheduling problems \citep{kelly2017crude,BRUNAUD2020106617}. Although this model facilitates temporal and phenomenological decomposition \citep{FRANZOI2024108678}, it is less suited for spatial decomposition, as spatial connectivity relationships are not explicitly defined by separate indices. Fig.~\ref{fig:supstrct} illustrates the key sets and variables required to properly define a process network using different superstructures.

From the perspective of process units, the process model represents the material transformation from raw inputs to finished products, directly influencing resource allocation and serving as the primary driver for optimizing production plans. Fixed-yield models, though computationally efficient, struggle to accurately capture the complex mechanistic properties of real processes. Conversely, nonlinear process models offer more rigorous simulations and higher profit potential but at the cost of increased computational perplexity \citep{MORO1998S1039,Siamizade2019}. To strike a balance, modeling efforts typically focus on high-throughput units, such as crude distillation units (CDUs) and fluid catalytic cracking (FCC) units. For CDUs, common modeling approaches include assigning fractions between CDU cut points (swing-cuts) \citep{zhang2001}, developing empirical relationships based on underlying process mechanisms \citep{li2007novel}, applying high-order polynomial approximations \citep{lopez2013}, and employing data-driven surrogate models \citep{li2016data}. Among these, the swing-cut formulation is widely adopted in industrial planning software due to its ease of implementation \citep{URIBERODRIGUEZ2020106883}.

Early attempts at solving refinery planning problems focused on eliminating bilinear terms introduced by pooling operations, typically through guessing initial values for unknown variables \citep{haverly1978studies} or applying Taylor expansions \citep{baker1985successive}. However, these methods often exhibit instability and produce weak relaxations. More advanced relaxation techniques, such as McCormick envelopes \citep{mccormick1976computability}, the reformulation-linearization technique (RLT) \citep{quesada1995global}, and the normalized multiparametric disaggregation technique (NMDT) \citep{castro2016normalized}, were developed to improve the quality of linearized formulations. Note that tighter piecewise linear relaxation models necessitate the inclusion of additional integer variables, thereby resulting in more complex mixed-integer linear programming (MILP) problems. These are usually solved using Branch \& Bound (B\&B) solvers \citep{GROSSMANN20122}. Global optimization methods for MINLP problems typically involve alternating between solving a relaxed dual problem and a heuristically derived projection of the original problem. The solutions provide upper and lower bounds within a search tree, with global optimality achieved when the difference between these bounds falls within a predefined tolerance. As problem size escalates, there is a growing interest on designing tailored solution algorithms that incorporate more efficient relaxation models \citep{CASTRO2021107459}, variable boundary tightening based on feasibility or optimality \citep{chen2022variable}, and spatial or temporal aggregation heuristics \citep{URIBERODRIGUEZ2023108229,YANG2023108121}.

Despite advances in computational techniques, the absence of a data-transparent benchmark that properly reflects real-world refinery operations remains a major obstacle to the further development of refinery planning methodologies \citep{PISTIKOPOULOS2021107252}. Clearly defined, reproducible, and realistic benchmark problems are crucial across all fields. Well-known benchmarks in Process Systems Engineering (PSE) include the Williams-Otto reactor for real-time optimization \citep{Williams1960} and the Tennessee Eastman process for process control \citep{DOWNS1993245}. To the best of the authors' knowledge, unlike established numerical optimization benchmark libraries such as MINLPLib \citep{Bussieck2003114} and MIPLIB \citep{miplib2017}, there are currently no refinery-petrochemical planning benchmarks that provide full access to model parameters and appropriately capture the intricacy of material networks and production requirements within such facilities. Although the pooling problem is one of the core problems in refinery planning and has a relatively well-developed collection of benchmark problems for algorithm validation \citep{TELES20093736,alfaki2013strong}, these benchmarks still fall short of capturing the full characteristics of the refinery planning problem. The refinery planning problem presents significant challenges due to the involvement of hundreds of materials with varying qualities and complex processing networks, alongside data privacy concerns. Many existing studies either simplify key aspects of the problem or lack sufficient data transparency, hindering researchers from fully validating their methods and performing comparative analysis. Table~\ref{tab:review} outlines representative refinery planning works from the past decade, evaluating each on modeling methodology, practical requirement implementation, and data accessibility. The lack of such benchmarks results in a significant disconnect between theoretical research and industrial applications.

\begin{table}[htb]
\setlength{\belowcaptionskip}{10pt}
\caption{Representative works on refinery planning}
\label{tab:review}
\resizebox{\textwidth}{!}{%
\begin{tabular}{lllllllll}
\hline
 &
   &
   &
   &
  \multicolumn{4}{c}{Practical requirement} &
   \\ \cline{5-8}
Reference &
  Network configuration &
  CDU &
  Process unit &
  \begin{tabular}[c]{@{}l@{}}Multi-batch\\ process\end{tabular} &
  \begin{tabular}[c]{@{}l@{}}Virtual\\ batch\end{tabular} &
  \begin{tabular}[c]{@{}l@{}}Operating\\ mode\end{tabular} &
  Inventory &
  Parameter accessability \\ \hline
\citep{castillo2017global} (2017) &
  \begin{tabular}[c]{@{}l@{}}Stream-based superstructure;\\ Simplified network\end{tabular} &
  Fixed-yield &
  Fixed-yield &
   &
   &
  \checkmark &
   &
  Fully accessible \\
\citep{URIBERODRIGUEZ2020106883} (2020) &
  \begin{tabular}[c]{@{}l@{}}Unit-stream-unit superstructure;\\ Comprehensive network\end{tabular} &
  Fixed-yield &
  Empirical correlation &
   &
   &
  \checkmark &
   &
  Partially accessible \\
\citep{LI2021107361} (2021) &
  \begin{tabular}[c]{@{}l@{}}Stream-based superstructure;\\ Simplified network\end{tabular} &
  Product trisection &
  Fixed-yield &
   &
   &
  \checkmark &
  Storage cost &
  Partially accessible \\
\citep{zhang2023enterprise} (2023) &
  \begin{tabular}[c]{@{}l@{}}Unit-stream-unit superstructure;\\ Comprehensive network\end{tabular} &
  Fixed-yield &
  Fixed-yield &
  \begin{tabular}[c]{@{}l@{}}\checkmark\\ \textit{(Scheduling)}\end{tabular} &
   &
  \checkmark &
  Storage cost &
  Not accessible \\
\citep{wang2024refinery} (2024) &
  \begin{tabular}[c]{@{}l@{}}Unit-stream-unit superstructure;\\ Simplified network\end{tabular} &
  Fixed-yield &
  Fixed-yield &
   &
   &
  \checkmark &
  Storage cost &
  Partially accessible \\
This work &
  \begin{tabular}[c]{@{}l@{}}Port-stream superstructure;\\ Comprehensive network\end{tabular} &
  Swing-cut &
  Delta-base and fixed-yield &
  \checkmark &
  \checkmark &
  \checkmark &
  \begin{tabular}[c]{@{}l@{}}Sophisticated stock\\ value accounting\end{tabular} &
  Fully accessible \\ \hline
\end{tabular}%
}
\end{table}

Building on the preceding discussion, a benchmark problem with strong applicability to engineering should possess the following characteristics: (1) a well-defined mathematical formulation and a comprehensive, accessible set of parameters to ensure reproducibility; (2) a good trade-off between model accuracy and computational tractability; (3) the capability to capture key constraints and operational requirements reflective of real-world production environments; and (4) broad generalizability, including aspects such as topological network configurations and the implementation of advanced solution methodologies. To develop such a benchmark, we first gathered production data from real-world refineries, analyzed production requirements across various scenarios, and established a mathematical formulation based on a novel port-stream hybrid superstructure. This superstructure leverages the strengths of both stream and port indexing, enabling efficient inventory configuration and spatial decomposition while offering flexibility for the implementation of algorithms, such as error-propagation-based methods, without compromising scalability. Utilizing real-world data, we then design three cases that encompass distinct scenario characteristics, including the scaling up in problem size, the incorporation of inventory-related integer variables, and the consideration of multi-period decision-making. The validity of the mathematical model is assessed through comparisons with actual refinery plans, allowing for iterative refinements. Finally, all parameters are systematically perturbed to ensure data confidentiality, and the modified dataset undergoes extensive testing to confirm the feasibility and robustness of the benchmark problem.

The principal contributions of this work can be concluded as follows:
\begin{enumerate}
    \item A production planning benchmark for real-world refinery-petrochemical complexes containing three cases is proposed, with all parameters accessible.
    \item A novel port-stream hybrid superstructure is formulated as the foundational framework for the mathematical model. The benefits of stream and port indexing are combined in this superstructure, which makes inventory configuration and spatial decomposition more efficient. It also offers flexibility in designing algorithms, including those based on error propagation, without compromising scalability.
    \item The presented benchmark formulation incorporates general practical production constraints and is designed for easy generalization to accommodate customized constraints in various scenarios.
    \item The performance of advanced commercial solvers is evaluated using the proposed benchmark, with a detailed analysis of their computational efficiency and solution quality. The impact of delta-base modeling for key process units is also assessed through a comparative feasibility study of linearly approximated solutions.
\end{enumerate}

The remainder of this paper is organized as follows. Section~\ref{sec:state} gives a brief overview of the processing routes in an integrated refinery-petrochemical plant. Section~\ref{sec:bm} describes the modeling framework and detailed mathematical formulation. Section~\ref{sec:case} demonstrates the computational results of the real-world case studies and discusses the significance and practical implications of the proposed benchmark problem. Finally, Section~\ref{sec:conclusion} offers concluding remarks.

\section{Problem Statement}
\label{sec:state}
Fig.~\ref{fig:flowchart} presents a typical flowchart of an integrated refinery-petrochemical complex. For clarity, the abbreviations of operation units and products are provided in the Nomenclature. Crude oil enters the complex via vessels or pipelines and undergoes fractionation in CDUs into various distillation cuts, including naphtha, kerosene, diesel, wax oil, and residual oil. The yields of these pseudo-components are derived from crude assay data. These fractions are then allocated to secondary processing units for deep conversion and upgrading based on their properties and desired product standards, with the goal of optimizing resource utilization and increasing value-added outputs. 

Straight-run gasoline, due to its low octane number, is usually processed in the continuous reformer (CRU), where it is reformed along with hydrocracking heavy naphtha. This process increases the aromatic content and octane number, generating high-value gasoline blending components. The CRU also produces hydrogen as a by-product, which serves as a crucial feedstock for subsequent hydrogenation units designed to improve the quality of other fractions. Aromatic components extracted from reformer products can be further fed into the chemical plant to produce downstream chemical products. Straight-run and hydrotreated wax oils are utilized as feedstocks for FCC units and hydrocracking (HDC) units. The FCC unit breaks down heavy oil into light gasoline, diesel and liquefied petroleum gas (LPG). The LPG is further processed in gas fractionation units (GFUs) to extract low-carbon olefins. Hydrocracking improves the quality of middle distillates through hydrogenation, yielding high-quality diesel and jet fuel while providing wax oil feedstock for the chemical plant. Residual oil undergoes further conversion through the delayed coking units (DCUs) to produce coker gas oil (CGO), coker diesel and petroleum coke. Both CGO and coker diesel require hydrogenation for quality enhancement before used as blending components.

The chemical production units are closely coupled with the refining operations, creating a mutually supportive and cyclic production network. The ethylene cracker is regarded as the core unit of the chemical plant, which relies on key feedstocks such as naphtha, LPG and hydrocracking tail oil supplied by the refinery. In return, the ethylene cracker provides critical by-products, including hydrogen and fuel gas, to support refinery operations, and some cracked gasoline may be recycled back to the refinery for further processing. Similarly, the aromatics complex (ARU) plays a vital role by extracting aromatics from reformer products for chemical production while returning by-products, such as light hydrocarbons, to the refinery for reuse as feedstock. The products of ethylene cracking are subsequently processed to manufacture essential chemicals, including polyethylene (PE), polypropylene (PP) and butadiene. This integrated framework underscores the importance of aligning feedstock characteristics with processing pathways to achieve maximum efficiency and product quality. While this overview outlines the key processing routes in an integrated refinery-petrochemical complex, the actual decision-making process is far more intricate. It involves complex and refined resource allocation strategies, particularly the selection of optimal processing routes for various materials. These considerations are encapsulated in the refinery planning problem, which can be formally described as follows:

\begin{figure}[htb]
    \centering
    \includegraphics[width=\textwidth]{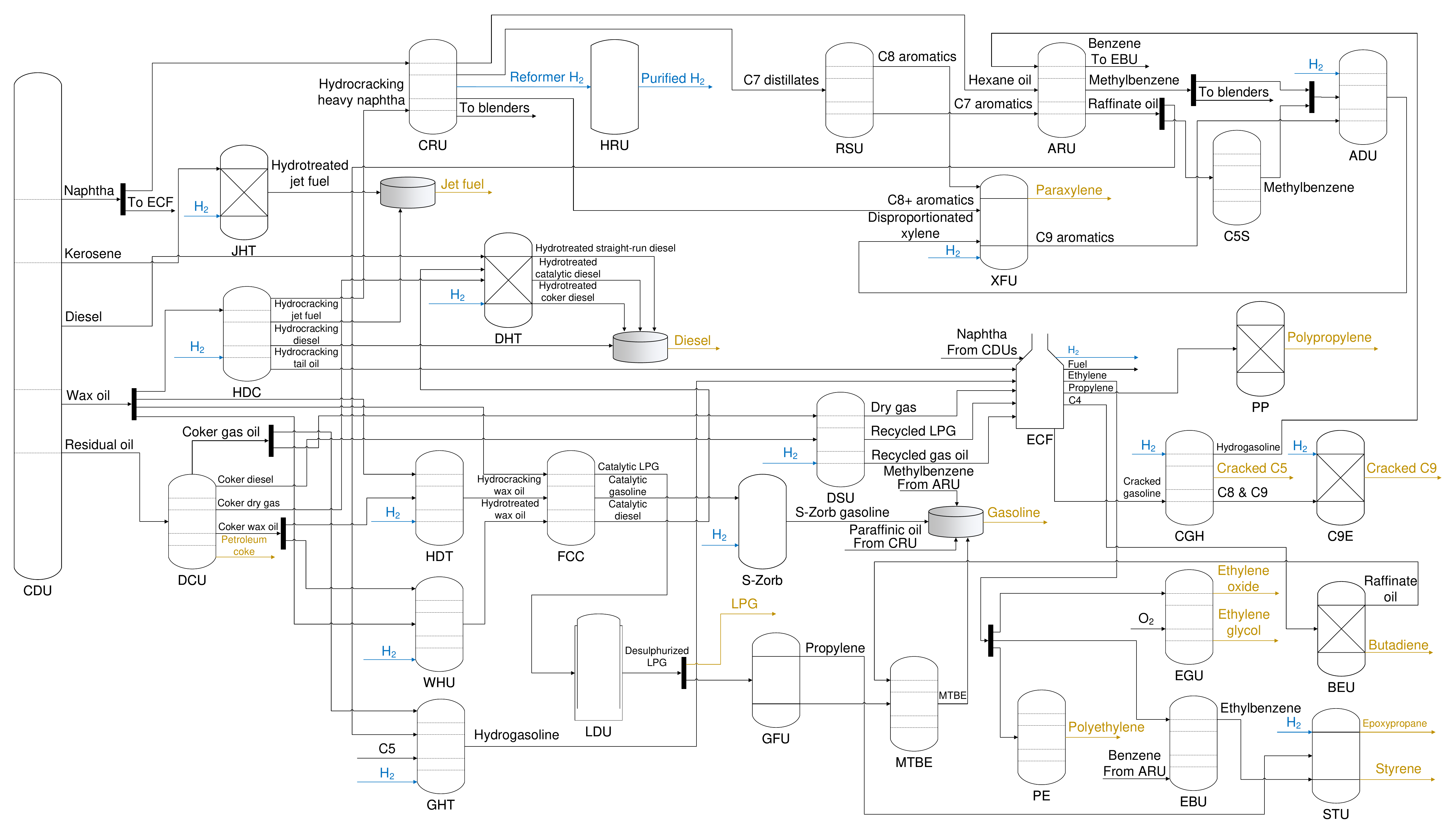}
    \caption{A simplified flowchart of an integrated refinery-petrochemical complex}
    \label{fig:flowchart}
\end{figure}

\textbf{Given:}
\begin{enumerate}
    \item The topology of the refinery, including the feed and output of individual process units.
    \item Crudes information, including their costs, properties, and product yields.
    \item Purchased raw materials information, including their costs, properties, and supply limits.
    \item Process units parameters, including their processing capacity limitations, feed quality specifications and product yields.
    \item Products information, including their costs, demands, and quality specifications.
    \item Inventory information, including initial levels, inventory cost factors, and whether a material can be stored.
\end{enumerate}

\textbf{Operation practices and assumptions:}
\begin{enumerate}
    \item The refinery topology is considered fixed, with all operations assumed to occur under steady-state conditions.
    \item The CDUs are modeled by the swing-cut model.
    \item The majority of secondary processing units are modeled using a fixed-yield approach, whereas the crucial units are represented using the delta-base modeling technique.
    \item All mixing operations are considered perfect mixing.
    \item Individual planning periods are long enough for one unit to handle multiple feed batches.
    \item Inventory profits result from the overproduction and subsequent storage of products, while economic losses occur when raw materials are withdrawn from inventory.
\end{enumerate}

\textbf{Determine:}
\begin{enumerate}
    \item The type and amount of crude oil processed during each time period.
    \item The throughput and feed composition of each process unit during each time period.
    \item The amount of purchased raw materials during each time period.
    \item The quantity and property of each product during each time period.
    \item The inventory profiles of concerned materials.
\end{enumerate}

The aim of multi-period refinery planning is to maximize total profit, which is calculated as the total economic value generated by all products, minus the costs associated with raw material procurement and material storage.

\section{Mathematical formulation}
\label{sec:bm}
\begin{figure}[htb]
    \centering
    \includegraphics[scale=0.8]{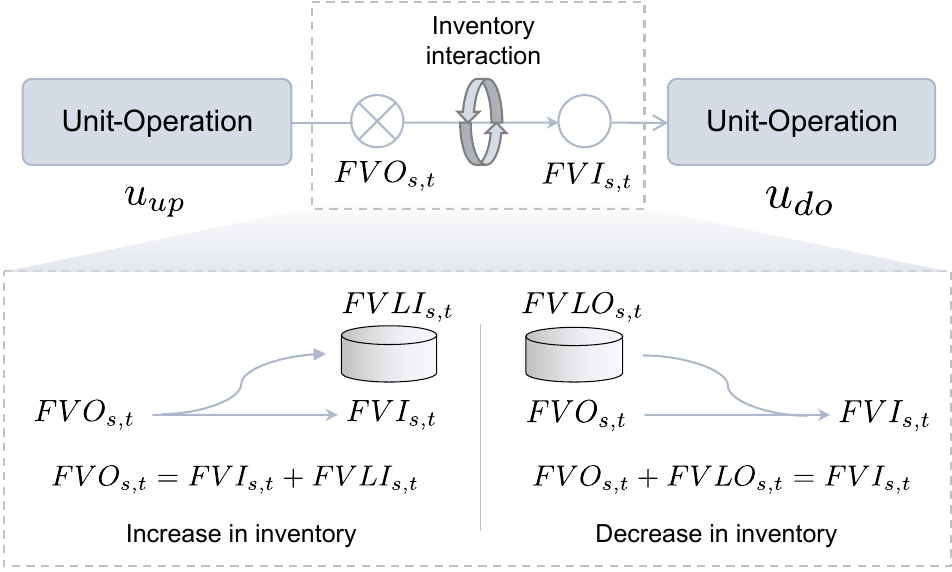}
    \caption{Illustration of port-stream superstructure}
    \label{fig:inv}
\end{figure}

The mathematical model in this paper is formulated based on a hybrid port-stream superstructure, where each flow variable is uniquely defined through a stream index as well as its location relative to the unit-operation. Specifically, the outlet mass flow variable $FVO_{s,t}$ and the inlet mass flow variable $FVI_{s,t}$ are introduced, where $s$ is the material stream index and $t$ represents the planning periods. This concept is illustrated in Fig.~\ref{fig:inv}. $FVLI_{s,t}$ and $FVLO_{s,t}$ represent the amount of stream $s$ added to or extracted from inventory during period $t$, respectively. In contrast to the UOPSS superstructure, which defines flows through the out-port of the upstream unit and the in-port of the downstream unit, the proposed superstructure explicitly defines distinct stream indices to facilitate spatial decomposition while maintaining consistency throughout the network in the presence of material inventories. This eliminates the additional effort required for storage tank modeling. Inventories can be configured by adjusting the bounds of the inventory interaction variables, $FVLI_{s,t}$ and $FVLO_{s,t}$, without changing the topology of the refinery-petrochemical complex.

This work considers five unit-operations: CDUs, process units, mixers, splitters, and blenders. Together, these operations encompass the essential processes within the refinery-petrochemical complex. Each unit operation exhibits distinct characteristics and is interconnected through the aforementioned port-stream superstructure, which ultimately constitutes a comprehensive mathematical model for production planning. The material balance constraints are first addressed, followed by a detailed discussion of the mathematical models governing each unit-operation individually. The sets, parameters, and variables referenced in this section are defined in the Nomenclature.

\subsection{Material balance}
\label{ssec:mb}
\begin{figure}[htb]
    \centering
    \includegraphics[scale=0.68]{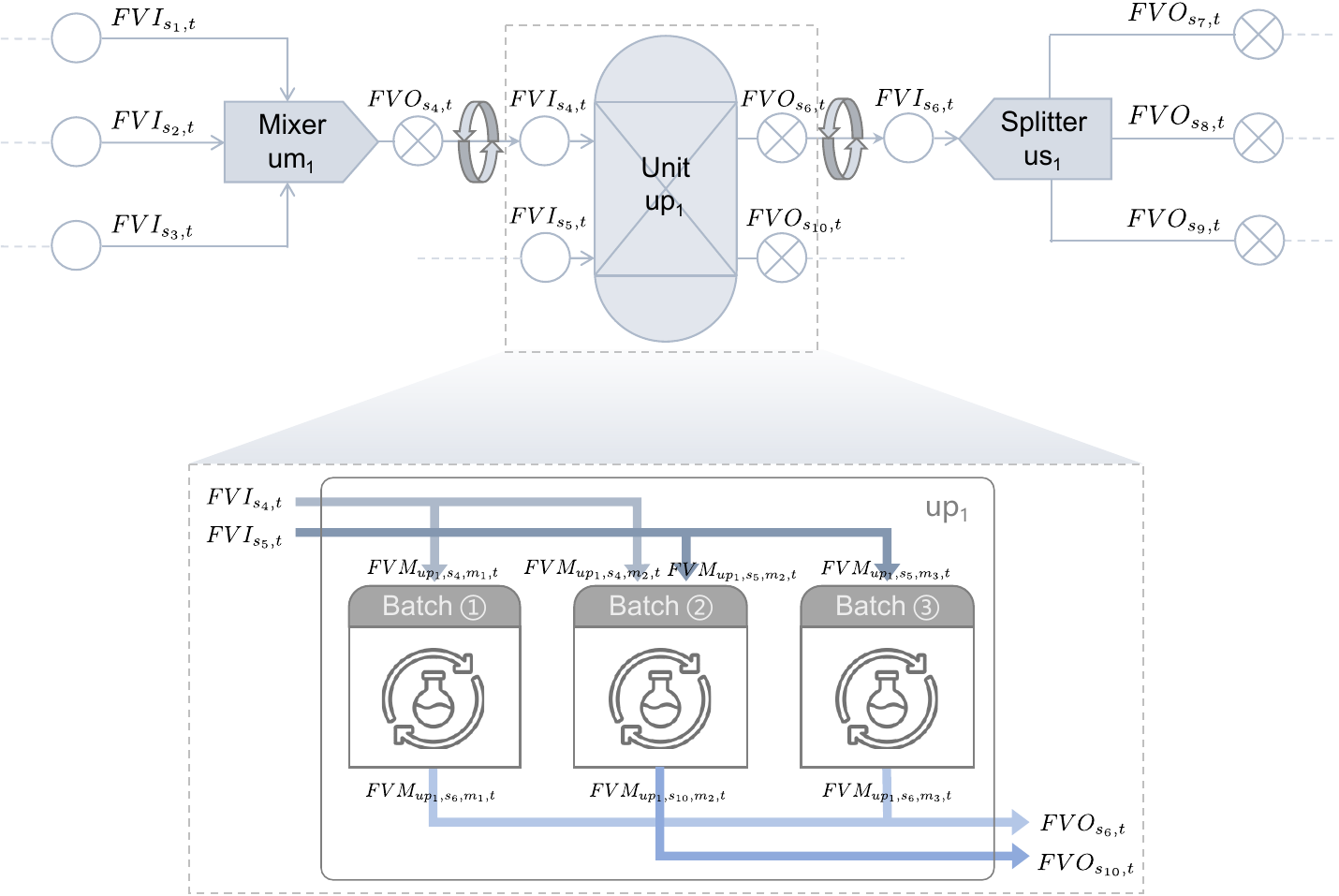}
    \caption{General unit model for batch material balance}
    \label{fig:batch}
\end{figure}

Unlike scheduling problems, planning problems tend to have longer planning horizon. Within a single period, a process unit may operate in multiple processing modes, requiring the planner to focus on the allocation of feed resources rather than on the detailed execution of specific operational schedules. Consequently, certain unit-operations can be modeled as batch processes, wherein the inlet and outlet flows represent the aggregated quantities of several processing batches (Fig.~\ref{fig:batch}).
\begin{subequations}\label{eq:batchsum}
    \begin{equation}
        FVI_{s,t}=\sum_{m:(u,m,s)\in IM}FVM_{u,m,s,t},\ \forall u\in U_{CDU}\bigcup U_{PF}\bigcup U_{PD}\bigcup U_{MIX},(u,s)\in IU
    \end{equation}
    \begin{equation}
        FVO_{s,t}=\sum_{m:(u,m,s)\in OM}FVM_{u,m,s,t},\ \forall u\in U_{CDU}\bigcup U_{PF}\bigcup U_{PD}\bigcup U_{MIX},(u,s)\in OU
    \end{equation}
\end{subequations}
where $m$ stands for different batches, and $u$ denotes the unit-operations. $(u,s)\in IU$ indicates that $s$ serves as the inlet stream of unit $u$, and $(u,m,s)\in IM$ states that $s$ is the inlet stream of unit $u$ in batch $m$. Similarly, $OU$ and $OM$ represent the corresponding outflow streams for units and batches, respectively. $FVM_{u,m,s,t}$ is the disaggregated mass flow of stream $s$ in batch $m$. Note that Eq.~\eqref{eq:batchsum} holds for all unit-operations that can be categorized as CDUs, process units (fixed-yield or delta-base) or mixers.

For mixers, the total amount of incoming material is equal to the total amount of outgoing material for each batch.
\begin{equation}
    \sum_{s:(u,m,s)\in IM}FVM_{u,m,s,t}=\sum_{s:(u,m,s)\in OM}FVM_{u,m,s,t},\ \forall m,t,u\in U_{MIX}
\end{equation}

For splitters and blenders, no multiple batches are considered. The material balance constraint is formulated as follows:
\begin{equation}
    \sum_{s:(u,s)\in IU}FVI_{s,t}=\sum_{s:(u,s)\in OU}FVO_{s,t},\ \forall t,u\in U_{SPL}\bigcup U_{BLD}
\end{equation}
where $U_{SPL}$ is the collection of all splitters, and $U_{BLD}$ is the collection of all blenders.

\subsection{CDU}
\label{ssec:cdu}
The swing-cut model is employed to represent CDUs. This formulation offers a good trade-off between precision and computational complexity, and is widely adopted in industrial applications \citep{khor2017petroleum}. It captures the dynamic nature of CDUs by facilitating the adjustment of product cutpoints in response to fluctuations in feed composition and operational conditions. Crude oil assays are represented by key product streams, known as heart cuts, with swing cuts enabling fine-tuned modifications to optimize yield distribution. Positioned between the main fractions, these swing cuts enhance flexibility within the distillation process. The output from the CDUs is calculated as follows:
\begin{equation}
\begin{split}
    FVM_{u,m,s,t}=\sum_{s':(u,m,s')\in IM}y_{u,m,s,s'}\cdot FVM_{u,m,s',t}\\+\sum_{\substack{s':(u,m,s')\in IM\\s'':(u,m,s,s'')\in SC}}y_{u,m,s'',s'}\cdot\phi_{u,m,s,s''}\cdot FVM_{u,m,s',t}\\-\sum_{\substack{s':(u,m,s')\in IM\\s'':(u,m,s'',s)\in SC}}y_{u,m,s,s'}\cdot\phi_{u,m,s'',s}\cdot FVM_{u,m,s',t},\\\forall t,u\in U_{CDU},(u,m,s)\in OM   
\end{split}
\end{equation}
where stream pairs in $SC$ indicate that cut fraction $s'$ can be merged into output stream $s$ of CDU $u$ in batch $m$. $\phi_{u,m,s,s'}$ represents the merge ratio, and $y_{u,m,s,s'}$ denotes the cut fraction yield.

In a similar manner, the properties of CDU products $FQ_{s,q,t}$ are calculated as follows:
\begin{subequations}
\label{eq:nonlinear1}
    \begin{equation}
    \begin{split}
        FVO_{s,t}/FQ_{s,q,t}=\sum_{m:(u,m,s)\in OM}(\sum_{s':(u,m,s')\in IM}y_{u,m,s,s'}\cdot FVM_{u,m,s',t}/FQ^{CUT}_{m,s,s',q}\\+\sum_{\substack{s':(u,m,s')\in IM\\s'':(u,m,s,s'')\in SC}}y_{u,m,s'',s'}\cdot\phi_{u,m,s,s''}\cdot FVM_{u,m,s',t}/FQ^{CUT}_{m,s'',s',q}\\-\sum_{\substack{s':(u,m,s')\in IM\\s'':(u,m,s'',s)\in SC}}y_{u,m,s,s'}\cdot\phi_{u,m,s'',s}\cdot FVM_{u,m,s',t}/FQ^{CUT}_{m,s,s',q}),\\\forall t,u\in U_{CDU},q=\text{SPG},(u,s)\in OU,(s,q)\in SQ,(s,q)\notin FIX
    \end{split}
    \end{equation}
    \begin{equation}
    \begin{split}
        FQ_{s,q',t}\cdot FVO_{s,t}/FQ_{s,q,t}\\=\sum_{m:(u,m,s)\in OM}(\sum_{s':(u,m,s')\in IM}y_{u,m,s,s'}\cdot FQ^{CUT}_{m,s,s',q'}\cdot FVM_{u,m,s',t}/FQ^{CUT}_{m,s,s',q}\\+\sum_{\substack{s':(u,m,s')\in IM\\s'':(u,m,s,s'')\in SC}}y_{u,m,s'',s'}\cdot\phi_{u,m,s,s''}\cdot FQ^{CUT}_{m,s'',s',q'}\cdot FVM_{u,m,s',t}/FQ^{CUT}_{m,s'',s',q}\\-\sum_{\substack{s':(u,m,s')\in IM\\s'':(u,m,s'',s)\in SC}}y_{u,m,s,s'}\cdot\phi_{u,m,s'',s}\cdot FQ^{CUT}_{m,s,s',q'}\cdot FVM_{u,m,s',t}/FQ^{CUT}_{m,s,s',q}),\\\forall t,u\in U_{CDU},q=\text{SPG},q'\in Q_V,(u,s)\in OU,(s,q')\in SQ,(s,q')\notin FIX
    \end{split}
    \end{equation}
    \begin{equation}
    \begin{split}
        FVO_{s,t}\cdot FQ_{s,q,t}=\sum_{m:(u,m,s)\in OM}(\sum_{s':(u,m,s')\in IM}y_{u,m,s,s'}\cdot FVM_{u,m,s',t}\cdot FQ^{CUT}_{m,s,s',q}\\+\sum_{\substack{s':(u,m,s')\in IM\\s'':(u,m,s,s'')\in SC}}y_{u,m,s'',s'}\cdot\phi_{u,m,s,s''}\cdot FVM_{u,m,s',t}\cdot FQ^{CUT}_{m,s'',s',q}\\-\sum_{\substack{s':(u,m,s')\in IM\\s'':(u,m,s'',s)\in SC}}y_{u,m,s,s'}\cdot\phi_{u,m,s'',s}\cdot FVM_{u,m,s',t}\cdot FQ^{CUT}_{m,s,s',q}),\\\forall t,u\in U_{CDU},q\in Q_W,(u,s)\in OU,(s,q)\in SQ,(s,q)\notin FIX
    \end{split}
    \end{equation}
\end{subequations}
where SPG is specific gravity, while $Q_V$ is the set of volume-based properties excluding SPG, and $Q_W$ is the set of weight-based properties. $SQ$ collects the concerned properties $q$ for various streams $s$. $(s,q)\in FIX$ states that certain property $q$ of stream $s$ is fixed during the optimization. Parameter $FQ^{CUT}_{m,s,s',q}$ denotes the property transfer ratio obtained from crude oil assays.

To ensure the smooth operation of the CDUs and maintain the specified quality of the cut fractions, the feed to the CDU must satisfy specific property constraints. These constraints are divided into two categories: unit-specific constraints for individual CDUs and aggregate constraints for the total amount of crude oil processed within a given time period.
\begin{subequations}
    \begin{equation}
    \begin{split}
        \underline{FQV_{u,m,q}}\cdot\sum_{\substack{s:(u,m,s)\in IM\\s:(s,q)\in SQ}}FVM_{u,m,s,t}/FQ^{CRD}_{m,s,q}\le\sum_{\substack{s:(u,m,s)\in IM\\s:(s,q)\in SQ}}FVM_{u,m,s,t}\\\le\overline{FQV_{u,m,q}}\cdot\sum_{\substack{s:(u,m,s)\in IM\\s:(s,q)\in SQ}}FVM_{u,m,s,t}/FQ^{CRD}_{m,s,q},\\\forall t,u\in U_{CDU},q=\text{SPG},(u,m,q)\in CDUMQ
    \end{split}
    \end{equation}
    \begin{equation}
    \begin{split}
        \underline{FQV_{u,m,q'}}\cdot\sum_{\substack{s:(u,m,s)\in IM\\s:(s,q)\in SQ}}FVM_{u,m,s,t}/FQ^{CRD}_{m,s,q}\le\sum_{\substack{s:(u,m,s)\in IM\\s:(s,q)\in SQ}}FQ^{CRD}_{m,s,q'}\cdot FVM_{u,m,s,t}/FQ^{CRD}_{m,s,q}\\\le\overline{FQV_{u,m,q'}}\cdot\sum_{\substack{s:(u,m,s)\in IM\\s:(s,q)\in SQ}}FVM_{u,m,s,t}/FQ^{CRD}_{m,s,q},\\\forall t,u\in U_{CDU},q=\text{SPG},q'\in Q_V,(u,m,q)\in CDUMQ
    \end{split}
    \end{equation}
    \begin{equation}
    \label{eq:cdufeed}
    \begin{split}
        \underline{FQV_{u,m,q}}\cdot\sum_{\substack{s:(u,m,s)\in IM\\s:(s,q)\in SQ}}FVM_{u,m,s,t}\le\sum_{\substack{s:(u,m,s)\in IM\\s:(s,q)\in SQ}}FQ^{CRD}_{m,s,q}\cdot FVM_{u,m,s,t}\\\le\overline{FQV_{u,m,q}}\cdot\sum_{\substack{s:(u,m,s)\in IM\\s:(s,q)\in SQ}}FVM_{u,m,s,t},\\\forall t,u\in U_{CDU},q\in Q_W,(u,m,q)\in CDUMQ
    \end{split}
    \end{equation}
\end{subequations}
where parameter $FQ^{CRD}_{m,s,q}$ designates crude properties, while $\underline{FQV_{u,m,q}}$ and $\overline{FQV_{u,m,q}}$ correspond to the minimum and maximum values of property $q$ for batch $m$ processed in CDU $u$, respectively. $CDUMQ$ specifies which properties are under control for each processed batches. Note that Eq.~\eqref{eq:cdufeed} also applies to the yield constraint for residue production.
\begin{subequations}
    \begin{equation}
        \begin{split}
            \underline{MFQ_q}\cdot\sum_{\substack{(u,m,s)\in IM\\u\in U_{CDU}}}FVM_{u,m,s,t}/FQ^{CRD}_{m,s,q}\le\sum_{\substack{(u,m,s)\in IM\\u\in U_{CDU}}}FVM_{u,m,s,t}\\\le\overline{MFQ_q}\cdot\sum_{\substack{(u,m,s)\in IM\\u\in U_{CDU}}}FVM_{u,m,s,t}/FQ^{CRD}_{m,s,q},\ \forall t,q\in\{\text{SPG}\}\cap CRU
        \end{split}
    \end{equation}
    \begin{equation}
        \begin{split}
            \underline{MFQ_q'}\cdot\sum_{\substack{(u,m,s)\in IM\\u\in U_{CDU}}}FVM_{u,m,s,t}/FQ^{CRD}_{m,s,q}\le\sum_{\substack{(u,m,s)\in IM\\u\in U_{CDU}}}FQ^{CRD}_{m,s,q'}\cdot FVM_{u,m,s,t}/FQ^{CRD}_{m,s,q}\\\le\overline{MFQ_q'}\cdot\sum_{\substack{(u,m,s)\in IM\\u\in U_{CDU}}}FVM_{u,m,s,t}/FQ^{CRD}_{m,s,q},\ \forall t,q=\text{SPG},q'\in Q_V\cap CRU
        \end{split}
    \end{equation}
    \begin{equation}
        \begin{split}
            \underline{MFQ_q}\cdot\sum_{\substack{(u,m,s)\in IM\\u\in U_{CDU}}}FVM_{u,m,s,t}\le\sum_{\substack{(u,m,s)\in IM\\u\in U_{CDU}}}FQ^{CRD}_{m,s,q}\cdot FVM_{u,m,s,t}\\\le\overline{MFQ_q}\cdot\sum_{\substack{(u,m,s)\in IM\\u\in U_{CDU}}}FVM_{u,m,s,t},\ \forall t,q\in Q_W\cap CRU
        \end{split}
    \end{equation}
\end{subequations}
where $\underline{MFQ_q}$ and $\overline{MFQ_q}$ correspond to the property bounds of the total crude oil processed in the system, and $CRU$ specifies which properties are under control.

\subsection{Process unit}
\label{ssec:pu}
In the existing literature on refinery planning, secondary processing units are typically modeled using a fixed-yield approach, which overlooks the impact of feed quality on product yields. While precise nonlinear models offer greater accuracy, they often introduce significant computational complexity. To strike a balance between realism and computational tractability, this study employs a delta-base technique to model the primary process units, such as FCC units, while conventional fixed-yield models are applied to the other units. This approach ensures more representative outcomes without imposing excessive computational burdens.

In this formulation, the base yield coefficient $\gamma_{u,m,s}$ is introduced for each stream involved in the reaction of process unit $u$, representing the quantitative relationship between inlet and outlet streams. Rather than defining a conversion rate tied to processing loads for each product, this coefficient enables more precise calculation of product yields across different batches and ensures proportional feed requirements for specific process units, such as hydrotreating units, thus enhancing both the flexibility and accuracy of yield estimations.
\begin{equation}
\begin{split}
    FVM_{u,m,s,t}=\gamma_{u,m,s}\cdot\sum_{s':(u,m,s')\in IM}FVM_{u,m,s',t}/\sum_{s':(u,m,s')\in IM}\gamma_{u,m,s'},\\\forall t,u\in U_{PF},(u,m,s)\in IM\bigcup OM
\end{split}
\end{equation}
\begin{equation}
\label{eq:pd}
    \begin{split}
        \Gamma_{u,m,s,t}-\gamma_{u,m,s}=\sum_{(s',q):(u,m,s',q)\in DBSQ}(FQ_{s',q,t}-B_{u,m,q})\cdot\delta_{u,m,s,q}/\Delta_{u,m,q},\\\forall t,u\in U_{PD},(u,m,s)\in IM\bigcup OM
    \end{split}
\end{equation}
\begin{equation}
\label{eq:nonlinear2}
    \begin{split}
        FVM_{u,m,s,t}=\Gamma_{u,m,s,t}\cdot\sum_{s':(u,m,s')\in IM}FVM_{u,m,s',t}/\sum_{s':(u,m,s')\in IM}\Gamma_{u,m,s',t},\\\forall t,u\in U_{PD},(u,m,s)\in IM\bigcup OM
    \end{split}
\end{equation}

Let $U_{PF}$ denote the set of fixed-yield process units and $U_{PD}$ the set of units modeled using the delta-base approach. The delta-base model employs delta-base vectors to account for yield variations caused by fluctuations in feed properties. These vectors linearize the nonlinear dynamics of refinery processes, allowing for more accurate modeling of unit performance. The base vector comprises the base feed property $B_{u,m,q}$ and the base yield coefficient $\gamma_{u,m,s}$, representing average unit performance. The delta vector depicts deviations through the change in feed property $\Delta_{u,m,q}$ and its corresponding effect on yield $\delta_{u,m,s,q}$. The adjusted yield coefficient $\Gamma_{u,m,s,t}$ is then computed using Eq.~\eqref{eq:pd}. $DBSQ$ specifies the pairs of streams and properties that affect product yields.

The properties of certain products are transmitted from the feed through the parameter $\alpha_{s,s',q}$.
\begin{equation}
    FQ_{s',q,t}=\alpha_{s,s',q}\cdot FQ_{s,q,t},\ \forall t,(s,q)\in SQ,(s,s',q)\in QT,(s',q)\notin FIX
\end{equation}
where $QT$ specifies the transfer of specific property $q$ between feed stream $s$ and output product $s'$.


\subsection{Mixer and splitter}
\label{sec:mns}
\begin{figure}[htb]
    \centering
    \includegraphics[scale=0.7]{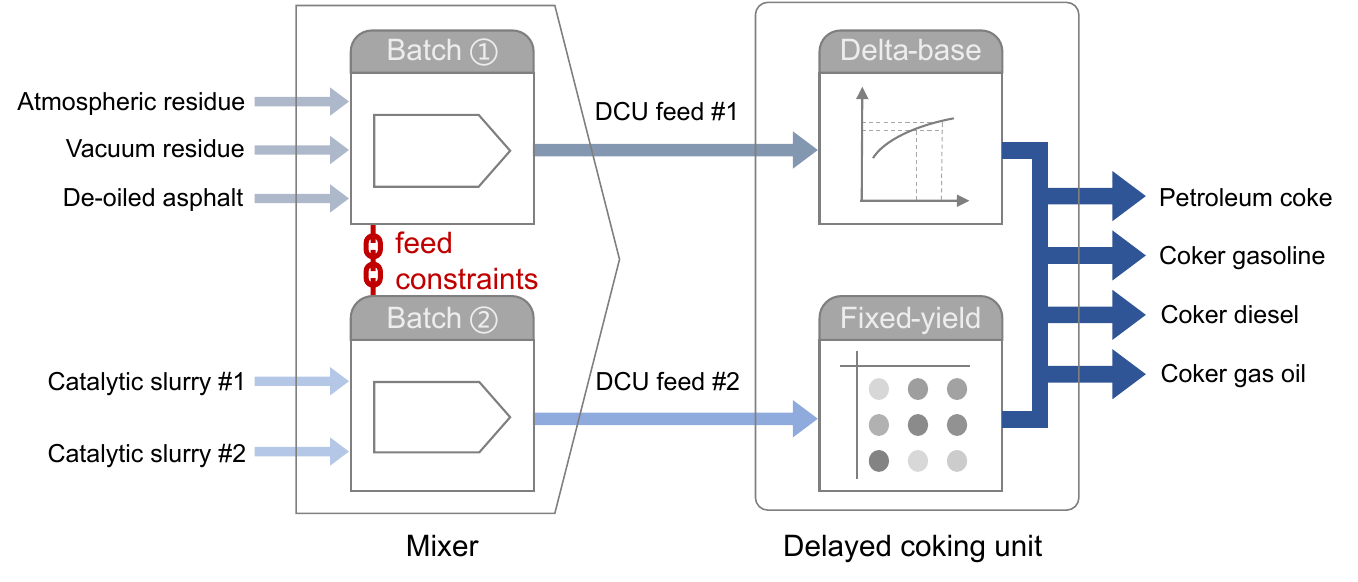}
    \caption{Illustration of the delayed coking process}
    \label{fig:vb}
\end{figure}

Intermediate products are pooled together via a mixer to serve as feed for downstream units, with the mixer formulated using the batch model, where each batch is assumed to have a single output. The rationale for employing a batch model rather than modeling each batch as an individual mixer lies in the need to account for the aggregate mixing properties of multiple feed batches in certain scenarios. This is demonstrated through the example of a DCU (Fig.~\ref{fig:vb}). In this scenario, the heavy oil batch consists of atmospheric residue, vacuum residue, and de-oiled asphalt. The coked products are calculated using the delta-base approach. Catalytic slurries from upstream units form the slurry batch, with the coked products calculated via the fixed-yield method. The property constraints of the DCU on the total feed interlink the two batches, making it inconvenient to model them separately.

Given that feed quality can influence product yields, it is essential to explicitly calculate the properties of the pooled feed streams. Variable $VM_{u,m,s,t}$ is introduced to denote the batch volumetric flow, while $V_{s,t}$ represents the total volumetric flow.
\begin{equation}
\label{eq:nonlinear3}
    VM_{u,m,s,t}\cdot FQ_{s,q,t}=FVM_{u,m,s,t},\ \forall t,u\in U_{MIX},q=\text{SPG},(s,q)\in SQ,(u,m,s)\in IM\bigcup OM
\end{equation}
\begin{subequations}
    \begin{equation}
        VM_{u,m,s,t}=\sum_{s':(u,m,s')\in IM}VM_{u,m,s',t},\ \forall t,u\in U_{MIX},(u,m,s)\in OM
    \end{equation}
    \begin{equation}
    \begin{split}
        VM_{u,m,s,t}\cdot FQ_{s,q,t}=\sum_{s':(u,m,s')\in IM}VM_{u,m,s',t}\cdot FQ_{s',q,t},\\\forall t,u\in U_{MIX},q\in Q_V,(s,q)\in SQ,(s,q)\notin FIX,(u,m,s)\in OM        
    \end{split}
    \end{equation}
    \begin{equation}
    \begin{split}
        FVM_{u,m,s,t}\cdot FQ_{s,q,t}=\sum_{s':(u,m,s')\in IM}FVM_{u,m,s',t}\cdot FQ_{s',q,t},\\\forall t,u\in U_{MIX},q\in Q_W,(s,q)\in SQ,(s,q)\notin FIX,(u,m,s)\in OM        
    \end{split}
    \end{equation}
\end{subequations}

Most feed property constraints can be enforced by applying them to the variable $FQ_{s,q,t}$. Howerver, for special cases, such as the one depicted in Fig.~\ref{fig:vb}, it is necessary to define a set of virtual batches related to the relevant properties, denoted as $VMQ$. For instance, the heavy oil batch and the slurry batch should be combined to form a new virtual batch, which is then subject to the property constraints as follows:
\begin{equation}
    V_{s,t}\cdot FQ_{s,q,t}=FVI_{s,t},\ \forall t,u\in U_{MIX},q=\text{SPG},(u,m,s,q)\in VMQ
\end{equation}
\begin{subequations}
\label{eq:nonlinear4}
    \begin{equation}
        \begin{split}
            \underline{FQV_{u,m,q}}\cdot\sum_{s:(u,m,s,q)\in VMQ}V_{s,t}\le\sum_{s:(u,m,s,q)\in VMQ}FVI_{s,t}\le\overline{FQV_{u,m,q}}\cdot\sum_{s:(u,m,s,q)\in VMQ}V_{s,t},\\\forall t,u\in U_{MIX},q=\text{SPG}
        \end{split}
    \end{equation}
    \begin{equation}
        \begin{split}
            \underline{FQV_{u,m,q}}\cdot\sum_{s:(u,m,s,q)\in VMQ}V_{s,t}\le\sum_{s:(u,m,s,q)\in VMQ}V_{s,t}\cdot FQ_{s,q,t}\\\le\overline{FQV_{u,m,q}}\cdot\sum_{s:(u,m,s,q)\in VMQ}V_{s,t},\ \forall t,u\in U_{MIX},q\in Q_V
        \end{split}
    \end{equation}
    \begin{equation}
        \begin{split}
            \underline{FQV_{u,m,q}}\cdot\sum_{s:(u,m,s,q)\in VMQ}FVI_{s,t}\le\sum_{s:(u,m,s,q)\in VMQ}FVI_{s,t}\cdot FQ_{s,q,t}\\\le\overline{FQV_{u,m,q}}\cdot\sum_{s:(u,m,s,q)\in VMQ}FVI_{s,t},\ \forall t,u\in U_{MIX},q\in Q_W
        \end{split}
    \end{equation}
\end{subequations}

In addition to general property constraints, specific components in the feed may also be subject to limitations. For example, the feed to an FCC unit is constrained by a restriction on the residue mixing ratio. Let $Q_P$ denote the property associated with the component ratio.
\begin{equation}
\label{eq:ratio}
\begin{split}
    \underline{FQV_{u,m,q}}\cdot\sum_{s:(u,s)\in IU}FVI_{s,t}\le\sum_{s:(u,m,s,q)\in VMQ}FVI_{s,t}\cdot w_{u,m,s,q}\le\overline{FQV_{u,m,q}}\cdot\sum_{s:(u,s)\in IU}FVI_{s,t},\\\forall t,u\in U_{MIX}\bigcup U_{PF}\bigcup U_{PD},q\in Q_P    
\end{split}
\end{equation}
where $w_{u,m,s,q}$ signifies the proportion of a specific component within stream $s$ for virtual batch $m$. Note that the virtual batches are defined in relation to each mixer in its entirety, and therefore, the aggregated flow variables $FVI_{s,t}$ and $V_{s,t}$ are used in the equations above. Eq.~\eqref{eq:ratio} is equally applicable in cases where process units are subject to compositional constraints on feedstock processing.

The splitter modeling is straightforward and requires only the assurance of material balance and property transfer.
\begin{equation}
    FQ_{s',q,t}=FQ_{s,q,t},\ \forall t,u\in U_{SPL},(u,s)\in IU,(u,s')\in OU,(s,q)\in SQ
\end{equation}

\subsection{Blender}
\label{ssec:bld}
Intermediate materials with diverse properties are blended to produce final products that meet specified quality criteria.
\begin{equation}
\label{eq:nonlinear5}
    V_{s,t}\cdot FQ_{s,q,t}=FVI_{s,t},\ \forall t,u\in U_{BLD},q=\text{SPG},(s,q)\in SQ,(u,s)\in IU
\end{equation}
\begin{subequations}
\label{eq:nonlinear6}
    \begin{equation}
        \underline{FQB_{u,q}}\cdot\sum_{s:(u,s)\in IU}V_{s,t}\le\sum_{s:(u,s)\in IU}FVI_{s,t}\le\overline{FQB_{u,q}}\cdot\sum_{s:(u,s)\in IU}V_{s,t},\ \forall t,u\in U_{BLD},q=\text{SPG}
    \end{equation}
    \begin{equation}
        \begin{split}
            \underline{FQB_{u,q}}\cdot\sum_{s:(u,s)\in IU}V_{s,t}\le\sum_{s:(u,s)\in IU}V_{s,t}\cdot FQ_{s,q,t}\le\overline{FQB_{u,q}}\cdot\sum_{s:(u,s)\in IU}V_{s,t},\\\forall t,u\in U_{BLD},q\in Q_V
        \end{split}
    \end{equation}
    \begin{equation}
        \begin{split}
            \underline{FQB_{u,q}}\cdot\sum_{s:(u,s)\in IU}FVI_{s,t}\le\sum_{s:(u,s)\in IU}FVI_{s,t}\cdot FQ_{s,q,t}\le\overline{FQB_{u,q}}\cdot\sum_{s:(u,s)\in IU}FVI_{s,t},\\\forall t,u\in U_{BLD},q\in Q_W
        \end{split}
    \end{equation}
\end{subequations}
where $\underline{FQB_{u,q}}$ and $\overline{FQB_{u,q}}$ indicate the bounds of the product specification. Note that each blender is associated with a single final product.

Certain blending components must be mixed in precise proportions to get the desired products, as described in Eq.~\eqref{eq:rb}. $u \in RB$ states that proportional blending occurs in blender $u$.
\begin{equation}
\label{eq:rb}
    FVI_{s',t}=\beta_{s,s'}\cdot FVO_{s,t},\ \forall t,u\in RB,(u,s)\in OU,(u,s')\in IU
\end{equation}

\subsection{Inventory management}
\label{ssec:inv}
The periods are connected by the variable $L_{s,t}$, which denotes the inventory level of stream $s$ at the end of period $t$.
\begin{equation}
    FVO_{s,t}+FVLO_{s,t}=FVI_{s,t}+FVLI_{s,t},\ \forall t,s
\end{equation}
\begin{subequations}
    \begin{equation}
        L_{s,t}=L^0_s+FVLI_{s,t}-FVLO_{s,t},\ \forall s,t=1
    \end{equation}
    \begin{equation}
        L_{s,t}=L_{s,t-1}+FVLI_{s,t}-FVLO_{s,t},\ \forall s,t>1
    \end{equation}
\end{subequations}
where $L^0_s$ is the initial inventory level of stream $s$.

Fluctuations in inventory levels affect the overall profit of the refinery-petrochemical complex; thus, a binary variable is introduced to ensure that these changes are properly recorded.
\begin{equation}
\label{eq:flaga}
    FVLO_{s,t}\le X_{s,t}\cdot\overline{L_{s,t}},\ \forall t,s
\end{equation}
\begin{equation}
\label{eq:flagb}
    FVLI_{s,t}\le(1-X_{s,t})\cdot\overline{L_{s,t}},\ \forall t,s
\end{equation}
where $X_{s,t}$ equals one if the inventory level of stream $s$ decreases during period $t$, and $\overline{L_{s,t}}$ represents the maximum inventory level of stream $s$ in each period. Eqs.~\eqref{eq:flaga} and~\eqref{eq:flagb} guarantee that at most one of the two variables $FVLO_{s,t}$ and $FVLI_{s,t}$ can be positive.

\subsection{Capacity control}
\label{ssec:cap}
To ensure the unit operates properly, the material throughput during a single period must be controlled. For most secondary processing units, regulating the total feed volume suffices. However, in certain units, the production levels of key products significantly influence resource allocation decisions in planning; thus, these specific product outputs serve as proxies for assessing the unit's processing capacity. To address this practical requirement, a capacity index $c$ is defined. Let $CAPIN$ be the set of capacities defined on input feeds, and $CAPOUT$ be the set of capacities defined on output products.
\begin{equation}
    \underline{FVC_{c,t}}\le\sum_{s:(c,s)\in CAPS}FVI_{s,t}\le\overline{FVC_{c,t}},\ \forall t,c\in CAPIN
\end{equation}
\begin{equation}
    \underline{FVC_{c,t}}\le\sum_{s:(c,s)\in CAPS}FVO_{s,t}\le\overline{FVC_{c,t}},\ \forall t,c\in CAPOUT
\end{equation}

\subsection{Hard bounds}
\label{ssec:hb}
The following constraints define the permissible value ranges for specific variables. Note that Eqs.~\eqref{eq:buy} and~\eqref{eq:sell} place restrictions on the quantity of raw materials procured and the volume of products sold, respectively.
\begin{equation}
    \underline{FQ_{s,q}}\le FQ_{s,q,t}\le\overline{FQ_{s,q}},\ \forall t,(s,q)\in SQ,(s,q)\notin FIX
\end{equation}
\begin{equation}
    FQ_{s,q,t}=FQ^0_{s,q},\ \forall t,(s,q)\in FIX
\end{equation}
\begin{equation}
    \underline{L_{s,t}}\le L_{s,t}\le\overline{L_{s,t}},\ \forall t,s
\end{equation}
\begin{equation}
\label{eq:buy}
    \underline{FV_{s,t}}\le FVO_{s,t}\le\overline{FV_{s,t}},\ \forall t,s\in S_M
\end{equation}
\begin{equation}
\label{eq:sell}
    \underline{FV_{s,t}}\le FVI_{s,t}\le\overline{FV_{s,t}},\ \forall t,s\in S_P
\end{equation}

\subsection{Objective function}
\label{ssec:obj}
\begin{figure}[htb]
    \centering
    \includegraphics[scale=0.55]{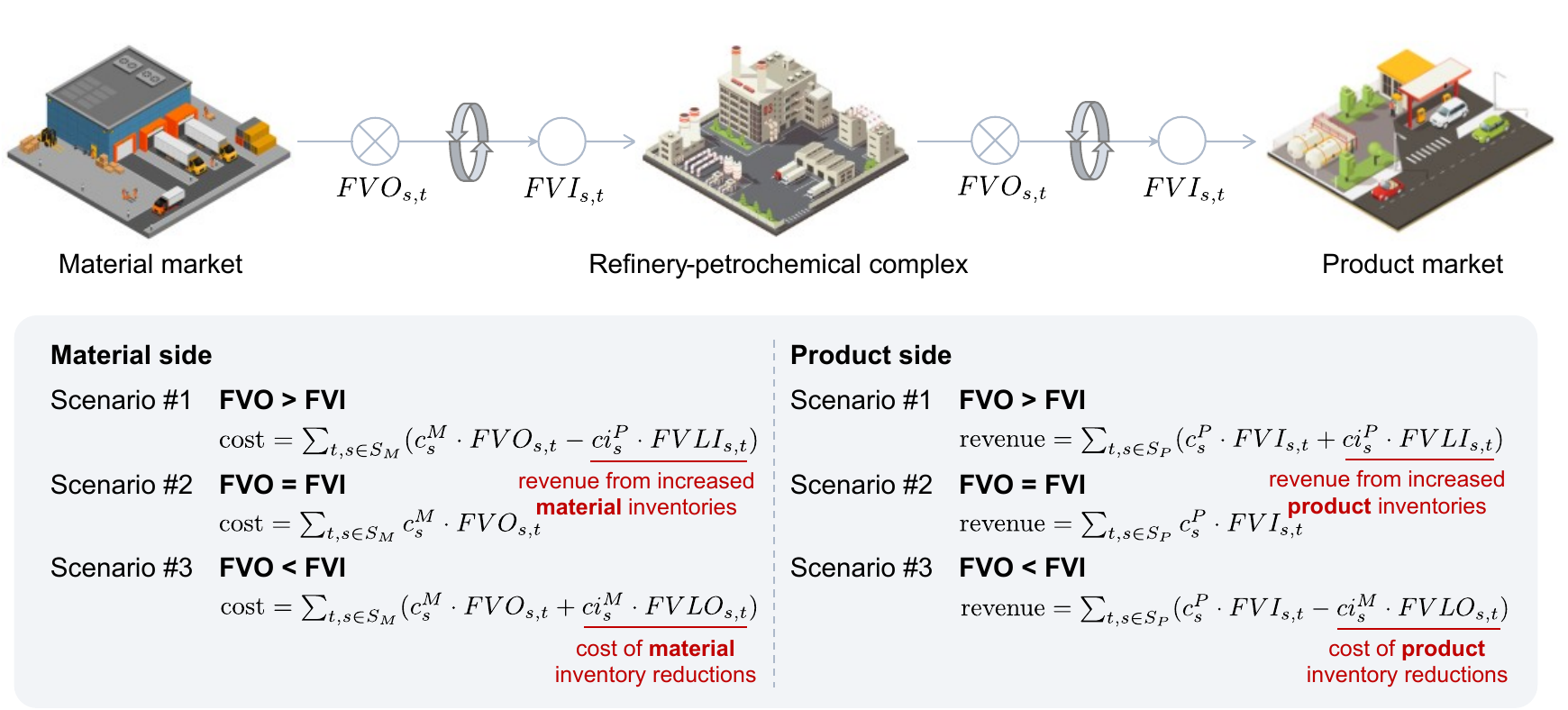}
    \caption{Illustration of profit calculation}
    \label{fig:obj}
\end{figure}
The objective of the mathematical model developed in this section is to maximize the profit, calculated by subtracting raw material procurement costs from product sales revenue plus the economic impact from inventory fluctuations. The detailed stock valuation method is depicted in Fig.~\ref{fig:obj}. On the material side, $FVO_{s,t}$ represents feedstock acquired directly from the market, which may be partially stored, with the remainder utilized in refinery operations. On the product side, $FVI_{s,t}$ denotes products entering the market, comprising both newly produced products and those previously stocked. With the associated economic coefficients defined, the objective function is formulated as follows:
\begin{equation}
\label{eq:obj}
    \begin{split}
        \text{Profit}=\sum_{t,s\in S_P}(c^P_s\cdot FVI_{s,t}+ci^P_s\cdot FVLI_{s,t}-ci^M_s\cdot FVLO_{s,t})\\-\sum_{t,s\in S_M}(c^M_s\cdot FVO_{s,t}+ci^M_s\cdot FVLO_{s,t}-ci^P_s\cdot FVLI_{s,t})
    \end{split}
\end{equation}

The proposed mathematical model of multi-period refinery-petrochemical planning problem can be stated as follows:
\begin{equation*}
    \begin{array}{lrl}
        \textbf{(RCP)} & \textbf{Max} & \text{Profit}\\
                        & \textit{s.t.} & \text{Eqs.~\eqref{eq:batchsum}-\eqref{eq:obj}}
    \end{array}
\end{equation*}

Note that \textbf{RCP} is an MINLP problem due to Eqs.~\eqref{eq:nonlinear1},~\eqref{eq:nonlinear2},~\eqref{eq:nonlinear3}-\eqref{eq:nonlinear4},~\eqref{eq:nonlinear5} and~\eqref{eq:nonlinear6}.

\section{Case Studies}
\label{sec:case}
To validate the wide applicability of the proposed mathematical formulation and enhance the diversity of the benchmark, three cases, each representing distinct characteristics of different operational scenarios, are derived from real-world refinery-petrochemical complexes. All model parameters have been systematically perturbed to preserve the confidentiality of proprietary data while retaining the essential characteristics of the problems. \textit{All sets and parameters are publicly accessible in \href{https://github.com/EMRPS/refinery-planning-benchmark}{\blue{\underline{EMRPS/refinery-planning-benchmark}}} for further reference.} 

Case 1 considers only the refinery site, excluding material exchanges between the refinery and the chemical plant. Additionally, no inventory is accounted for across any streams, resulting in a smaller-scale Nonlinear Programming (NLP) problem optimized in continuous space. Case 2 expands the problem boundary to incorporate material inter-supply between the refinery and chemical site while introducing inventories for raw materials, products, and intermediates. Binary indicators are employed to capture inventory fluctuations and assess their impact on economic objectives. While Cases 1 and 2 are single-period instances, Case 3 extends the planning horizon to three periods within a fully integrated refinery-petrochemical complex, enabling the evaluation of long-term decision-making effects.

Specifically, Case 1 describes a single-period refinery operation that involves two CDUs, which process suitable crude oil selected from twenty-five available crude types. The distillation fractions undergo further processing through twenty-five secondary units, including FCC, CRU, and DCU, resulting in the production of twenty-four distinct petrochemical products. Key fuel commodities, such as gasoline and diesel with varying specifications, are obtained by blending multiple intermediate components. In order to better capture the nonlinear relationship between feed properties and output yields, five critical process units are modeled using the delta-base technique. Furthermore, the model tracks a range of key properties, such as specific gravity, sulfur content, and research octane number, to ensure that the product meets its criteria. No inventory is considered for any material in this case. Variations in crude oil slates lead to differences in distillation fractions, including disparities in flow rates and material properties. These fractions serve as the main feedstock for downstream secondary processing units and directly influence operating modes, material flow directions, and product yields. For example, high-quality light gas oil (LGO) with low sulfur content and high aromatic concentration is preferentially routed to FCC units or CRUs to maximize value-added products. In contrast, LGO with higher impurity levels is typically directed toward hydrorefining or straight blending into diesel. Such choices are further affected by unit capacity constraints and market demand for specific products. Case 2 illustrates another single-period problem in which three CDUs process crude oil slates selected from fifteen available crude types. This instance represents a refinery complex integrated with a chemical site and comprises fifty-one secondary processing units, including eleven units modeled using the delta-base technique. The entire process produces forty-four distinct petrochemical products. Inventories are considered for raw materials, products and intermediates. Case 3 is a three-period extension of Case 2, with some parameters slightly different.

The model statics for studied cases are presented in Table~\ref{tab:modelstat}. All problems are implemented in GAMS 45.5.0 on a desktop equipped with an Inter(R) Core(TM) i7-8700 CPU 3.20 GHz and 16 GB of RAM. ANTIGONE and BARON are employed as global optimization solvers. Termination criteria are set to a relative tolerance of 0.01\% or a maximum CPU time of 18,000 seconds.

\begin{table}[htb]
\centering
\setlength{\belowcaptionskip}{10pt}
\caption{Model statistics of studied cases}
\label{tab:modelstat}
\resizebox{0.85\textwidth}{!}{%
\begin{tabular}{lllll}
\hline
Case & Total variables & Binary variables & Total constraints & Nonlinear elements \\ \hline
1    & 3573            & 0                & 3428              & 2082               \\
2    & 7157            & 56               & 8156              & 4294               \\
3    & 21469           & 168              & 24466             & 12882              \\ \hline
\end{tabular}%
}
\end{table}

In the following subsections, we first present the performance of the commercial MINLP solver solutions for the three cases introduced previously. This is followed by a detailed analysis of the necessity for nonlinear characterization of critical process units, as demonstrated through an ablation experiment conducted on Case 2. Finally, we provide a brief discussion of potential future research directions that could benefit from the benchmark constructed in this work.

\subsection{Computational results}
\label{ssec:results}
Table~\ref{tab:results} summarizes the computational performance of the studied cases using the commercial MINLP global solvers ANTIGONE and BARON. For all three cases, neither solver achieves a global optimum within the given time, highlighting the inefficiency of existing solvers in addressing industrial-scale refinery planning problems. Notably, ANTIGONE fails to identify any feasible solution for Cases 2 and 3, prematurely terminating before reaching the maximum runtime and providing only estimated upper bounds. The solver does not provide precise bound information when variable bounds cannot be not deterministically inferred. ANTIGONE mistakenly claims these cases are infeasible problems, whereas BARON manages to obtain feasible solutions. The poor performance of ANTIGONE may stem from its preprocessing strategy, which involves reformulating user-defined MINLP problems before applying linear relaxations \citep{misener2014antigone}. The reformulation assumptions may be incompatible with the proposed formulation. This can affect the structure of the search tree, which ultimately causes all nodes to be explored without feasible solution found. As for the simplest case, both solvers identify the same solution, although BARON reports a tighter upper bound.

\begin{table}[htb]
\centering
\setlength{\belowcaptionskip}{10pt}
\caption{Computational performance of studied cases}
\label{tab:results}
\resizebox{0.75\textwidth}{!}{%
\begin{tabular}{lllll}
\hline
Solver   &              & Case 1     & Case 2     & Case 3      \\ \hline
ANTIGONE & solution     & 34,167,968 & N/A        & N/A         \\
         & upper bound  & 41,582,417 & 76,070,000 & 133,700,000 \\
         & opt gap (\%) & 17.83      & N/A        & N/A         \\
         & CPU time (s) & 18000      & 225        & 396         \\
BARON    & solution     & 34,167,968 & 67,303,190 & 125,250,466 \\
         & upper bound  & 37,910,847 & 76,374,974 & 135,294,530 \\
         & opt gap (\%) & 9.87       & 11.88      & 7.42        \\
         & CPU time (s) & 18000      & 18000      & 18000       \\ \hline
\end{tabular}%
}
\end{table}

\begin{figure}[htb]
    \centering
    \includegraphics[scale=0.52]{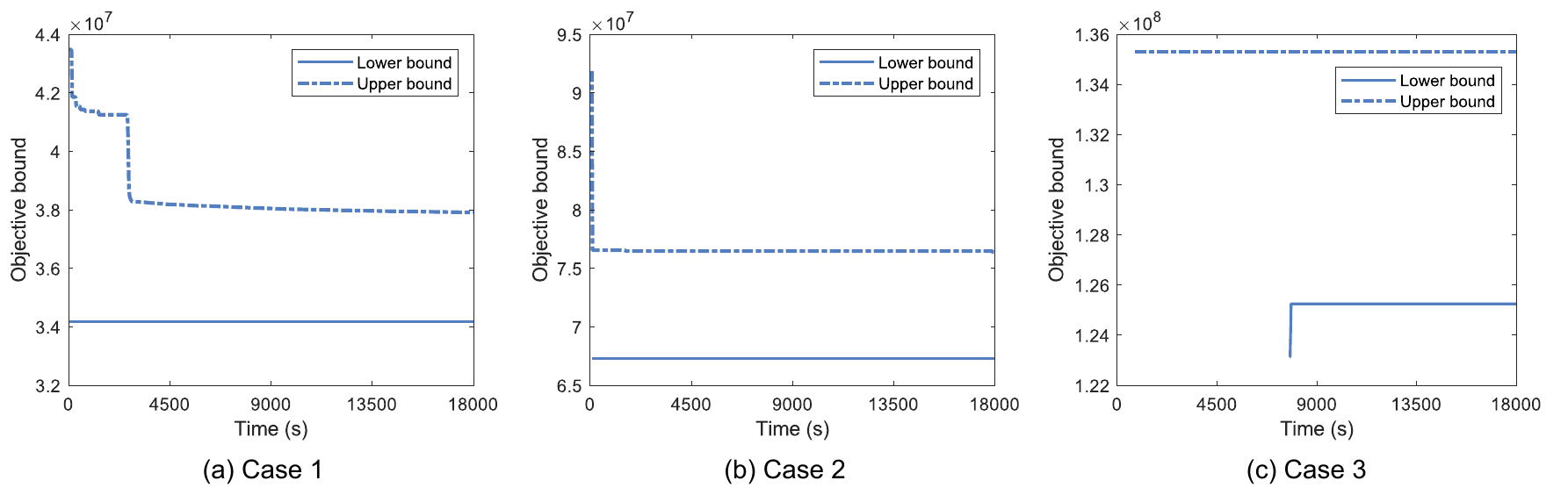}
    \caption{Objective bounds of studied cases}
    \label{fig:cvg}
\end{figure}

The computational challenges arise mainly from the nonconvexity introduced by binary variables, pooling constraints, and the delta-base formulation. This complexity is further exacerbated by the strong coupling of variables inherent in the extended material flow chains characteristic of petrochemical processes. Fig.~\ref{fig:cvg} plots the bound information returned by BARON with respect to the solution time for a better understanding of the solution process. Among the three cases, only Case 1 exhibits a noticeable continuous improvement in the upper bound throughout the entire solution procedure. This behavior can be attributed to the absence of binary variables in this particular instance. In contrast, the upper bounds for Cases 2 and 3 barely change following an initial improvement phase. This stagnation underscores the importance of developing tighter relaxations and incorporating stronger cutting planes to enhance computational performance. With regard to the lower bounds, BARON efficiently identifies a feasible solution for Cases 1 and 2 early in the computation process but fails to achieve any further improvement over time. For the multi-period Case 3, BARON requires approximately 2.2 hours to obtain the first feasible solution. These observations highlight BARON's inefficiency in exploring and pruning the branch-and-bound tree, which restricts its ability to discover better solutions for complex problems.

\subsection{Key process unit representation}
\label{ssec:db}
As discussed in Section~\ref{ssec:results}, the nonconvexity inherent in the model presents substantial computational challenges. This complexity is further amplified by the incorporation of the delta-base formulation for key process units. The feed properties of secondary processing units often require upstream bilinear pooling for accurate determination, rendering the product yield equations intrinsically nonlinear (Eq.~\eqref{eq:nonlinear2}). This nonlinearity directly impacts the distribution of product yields, which, in turn, influences resource allocation and property calculations for downstream units. Despite the decline in computational performance, accurate yield prediction plays a critical role in ensuring the feasibility and optimality of decision-making processes. While there are existing publications exploring the effects of nonlinear modeling of process units, these analyses predominantly focus on the comparison of economic outcomes \citep{Siamizade2019,DEMIRHAN2020107007}. In this section, a control version of Case 2 is solved, wherein all process units are modeled using a fixed-yield approach based on real industrial data. Examination of the resulting solutions reveals four scenarios where linear simplifications introduce errors that compromise feasibility. Note that this study represents the first attempt to analyze the feasibility of linearly approximated solutions within the context of an industrial case.

\begin{table}[htb]
\centering
\setlength{\belowcaptionskip}{10pt}
\caption{With and without delta-base formulation statistics}
\label{tab:db_cp}
\resizebox{0.75\textwidth}{!}{%
\begin{tabular}{llll}
\hline
\multicolumn{2}{l}{Model statistics}   & With delta-base & Without delta-base \\ \hline
\multicolumn{2}{l}{Total variables}    & 7157            & 6976               \\
\multicolumn{2}{l}{Total constraints}  & 8156            & 7957               \\
\multicolumn{2}{l}{Nonlinear elements} & 4294            & 3556               \\ \hline
\multicolumn{2}{l}{Solver statistics}  & With delta-base & Without delta-base \\ \hline
ANTIGONE          & solution           & N/A             & 66,990,616         \\
                  & upper bound        & 76,070,000      & 74,439,973         \\
                  & opt gap (\%)       & N/A             & 10.01              \\
                  & CPU time (s)       & 225             & 898                \\
BARON             & solution           & 67,303,190      & 66,517,916         \\
                  & upper bound        & 76,374,974      & 74,532,208         \\
                  & opt gap (\%)       & 11.88           & 10.75              \\
                  & CPU time (s)       & 18000           & 18000              \\ \hline
\end{tabular}%
}
\end{table}

Table~\ref{tab:db_cp} displays the model statistics and solution outcomes. Upon removing the delta-base formulation, ANTIGONE outperforms BARON, delivering a better solution. However, the search process still terminates prematurely, as all nodes in the branch-and-bound tree have been explored. Compared to the model incorporating the delta-base formulation, the objective value decreases. Moreover, certain decision variables gravitate toward the boundaries of the feasible domain, driven by the objective of profit maximization. These decisions are highly sensitive to model accuracy, with small deviations potentially rendering the production plan infeasible. This is particularly problematic because refinery planning outputs often serve as execution targets for lower-level scheduling systems, where production operations are scheduled sequentially in shorter time intervals. Unrealistic production plans can lead resource shortages, load overruns, and substandard products during subsequent scheduling stages. These feasibility challenges will be specified below in relation to the process in Case 2.

\textbf{The feed ratio for the process unit fails to comply with the specified requirements.} 
A representative example involves a series of reactions requiring hydrogen participation. For instance, coker gas oils (CGO) from two DCUs are combined as feed for downstream hydrocracking and hydrotreating (HDT) unit and wax oil hydrotreating unit (WHU). The yield of CGO is influenced by feed properties such as specific gravity, sulfur content, and Conradson carbon residue. This dependency results in an actual CGO yield exceeding the solution's predicted values (Table~\ref{tab:db_dcu}). Concurrently, the WHU operates at its maximum capacity, leaving the surplus CGO to be redirected to the HDT. However, the HDT lacks sufficient hydrogen, which subsequently affects the quality of hydrotreated wax oil. Note that the calibrated mass flow values in Table~\ref{tab:db_dcu} are computed using Eqs.~\eqref{eq:pd} and~\eqref{eq:nonlinear2}, wherein the feed property $FQ_{s',q,t}$ is derived from the solution of the fixed-yield model.

\begin{table}[htb]
\centering
\setlength{\belowcaptionskip}{10pt}
\caption{Yield predictions for DCUs in Case 2}
\label{tab:db_dcu}
\resizebox{0.75\textwidth}{!}{%
\begin{tabular}{llll}
\hline
Feed properties    & Specific gravity    & Sulfur content (\%) & Concarbon (\%) \\ \hline
\#1 DCU Feed       & 0.984078            & 3                   & 18             \\
\#2 DCU Feed       & 0.993018            & 1                   & 18             \\ \hline
\multicolumn{2}{l}{Product massflow (t)} & Fixed yield         & Calibration\\ \hline
\#1 DCU            & petroleum coke      & 1,354.7             & 1,250.4        \\
                   & coker gas oil       & 965.8               & 984.8          \\
                   & coker diesel        & 948.3               & 981.3          \\
                   & coker gasoline      & 751.6               & 801.5          \\
\#2 DCU            & petroleum coke      & 904.7               & 904.5          \\
                   & coker gas oil       & 932.0               & 932.0          \\
                   & coker diesel        & 682.8               & 682.5          \\
                   & coker gasoline      & 641.8               & 641.3          \\ \hline
\end{tabular}%
}
\end{table}

\textbf{The throughput of process unit falls outside the desirable range.} Yield variations are particularly pronounced in high-throughput units such as the FCC units. In Case 2, two FCC units process heavy oil and wax oil through catalytic cracking. The produced catalytic gasoline is subsequently sent to the S-zorb unit for desulfurization and refining. Due to the high profitability of gasoline products, the S-zorb unit operates at its maximum capacity. However, the yield of catalytic gasoline is influenced by feed properties, including specific gravity, sulfur content, Conradson carbon residue, and total nitrogen. These factors result in elevated yields that exceed the processing capacity of the downstream S-Zorb unit (Table~\ref{tab:db_fcc}).

\begin{table}[htb]
\centering
\setlength{\belowcaptionskip}{10pt}
\caption{Yield predictions for FCCs in Case 2}
\label{tab:db_fcc}
\resizebox{0.95\textwidth}{!}{%
\begin{tabular}{lllll}
\hline
Feed properties & Specific gravity         & Sulfur content (\%)        & Concarbon (\%) & Total nitrogen (ppm) \\ \hline
\#1 FCC Feed          & 0.893083                         & 0.7                & 3.39444     & 1500       \\
\#2 FCC Feed          & 0.85                             & 0.3                & 1.10333     & 1500       \\ \hline
\multicolumn{2}{l}{Product massflow (t)}                 &                    & Fixed yield & Calibration \\ \hline
\#1 FCC               & \multicolumn{2}{l}{catalytic gasoline}                & 1,463.6     & 1,498.0    \\
                      & \multicolumn{2}{l}{catalytic liquefied petroleum gas} & 771.5       & 805.0      \\
                      & \multicolumn{2}{l}{catalytic diesel}                  & 761.0       & 721.6      \\
                      & \multicolumn{2}{l}{catalytic slurry oil}              & 263.3       & 245.4      \\
\#2 FCC               & \multicolumn{2}{l}{catalytic gasoline}                & 2,865.2     & 2,880.3    \\
                & \multicolumn{2}{l}{catalytic liquefied petroleum gas} & 1,370.9        & 1,378.2              \\
                      & \multicolumn{2}{l}{catalytic diesel}                  & 1,381.4     & 1,376.8    \\
                      & \multicolumn{2}{l}{catalytic slurry oil}              & 337.4       & 330.4      \\ \hline
\end{tabular}%
}
\end{table}

\textbf{The market demand for the product cannot be fulfilled.} As previously showed in Table~\ref{tab:db_fcc}, the variations in feedstock properties will lead to an increase yield of catalytic gasoline from the FCC units, accompanied by a reduction in the yield of catalytic diesel. In Case 2, the solver prioritized meeting the lower bound of light fuel oil demand to capitalize on the higher profitability of other products. Note that catalytic diesel constitutes a significant component of light fuel oil. When the actual yield deviates from the model results, this segment of production will no longer meet the market demand for orders.

\textbf{The property of the blended product fails to meet the required standards.}
Inspection of the results of the model without delta-base formulation reveals that the aromatic content of \#92 gasoline reaches its upper limit (Table~\ref{tab:db_w92}). This is primarily due to the inclusion of components with high aromatic content. In order to meet the national standard, high-octane components, such as methyl tertiary-butyl ether (MTBE) and aromatics, need to be added during the gasoline blending process. MTBE is produced from methanol derived from catalytic liquefied petroleum gas (LPG) via GFUs or must be purchased when refinery methanol circulation is insufficient, incurring high costs. Aromatics, convertible from lower-carbon alkanes in the CRU, enhance the utilization of low-value alkanes and improve gasoline quality, making them essential for blending. However, excessive aromatic content leads to harmful emissions and must be strictly controlled. The production of these key components is sensitive to feedstock properties: catalytic LPG originates from FCCs, and aromatic-rich distillates are produced in CRUs. In this case, while the actual yields of MTBE and paraffinic oils exceed the fix-based predicted values, ensuring that the aromatic content of the blended product remained within permissible limits, the rationale outlined above could lead to undesired product properties under alternative resource allocation decisions. For instance, this could occur if the actual MTBE yield was lower than the predicted value.

\begin{table}[htb]
\centering
\setlength{\belowcaptionskip}{10pt}
\caption{\#92 gasoline property criteria and its blending components in Case 2}
\label{tab:db_w92}
\resizebox{\textwidth}{!}{%
\begin{tabular}{lllll}
\hline
Properties    & Specific gravity           & Sulfur content (\%) & Research octane number & Aromatic content (\%) \\ \hline
Fixed-yield solution              & 0.725001 & 0.00044 & 96.7924 & 34 \\
Specification & $\ge 0.72$ and $\le 0.775$ & $\le 0.0009$        & $\ge 92.45$            & $\le 34$              \\ \hline
Components    & S-Zorb gasoline            & Methylbenzene       & Paraffinic oil         & MTBE                  \\ \hline
Concentration (\%)    & 56       & 21      & 18      & 5  \\
Aromatic content (\%) & 22.245   & 100     & 0       & 0  \\ \hline
\end{tabular}%
}
\end{table}

The preceding analysis highlights the critical influence of yield predictions for key process units on the feasibility of production plans. The delta-base approach offers a well-balanced solution, bridging the gap between oversimplified fixed-yield models and computationally intensive nonlinear mechanistic models. This method serves as an effective tool for enhancing planning accuracy in industrial applications.

\subsection{Discussion}
\label{ssec:discussion}
This paper introduces the first open-source, demand-driven benchmark problem for integrated refinery-petrochemical complexes at an industrial scale. Substantial effort has been devoted to data collection and processing, as well as model development and validation. The authors anticipate that this benchmark will serve as a valuable resource for advancing the following essential subjects within the field of PSE:

(1) Development of efficient solution algorithms, with a particular focus on addressing bilinear and trilinear terms within process networks;

(2) Exploration of temporal and spatial decomposition mechanisms, including the identification of key process network clusters and the adaptive reconfiguration of multi-period planning frameworks;

(3) Characterization of material value and implementation of tiered pricing strategies across the production chain to investigate the dynamic interactions between supply and demand propagation;

(4) Investigation of methods for managing uncertainty, encompassing both precautionary planning approaches and rapid response mechanisms for addressing sudden disturbances; and

(5) Research on diagnostic traceability for infeasibility, leveraging network topology and solver numerical information to systematically identify the sources of infeasibility.

\section{Conclusion}
\label{sec:conclusion}
The pursuit of achieving digital intelligence transformation and carbon neutrality has set higher standards for integrated refinery-petrochemical enterprises. Refinery planning is a critical upstream step in the decision chain. Its associated technological advancements shape both the market competitiveness of individual enterprises and the broader trajectory of industrial development.  However, existing theoretical research on refinery planning lacks a reproducible benchmark problem with clear constraints, transparent parameters, and practical relevance. This paper addresses this gap by introducing a benchmark based on a novel port-stream hybrid superstructure, which effectively represents the refinery topology network. This approach enables seamless modeling of material stream inventories and facilitates modular construction of the process network. All process operations are systematically categorized into five classes according to their functions, namely, CDUs, process units, mixers, splitters, and blenders. They are represented by distinct sets of equations. To more accurately reflect actual industrial processes, the model employs the delta-base method for key secondary processing units, capturing the nonlinear relationships between product yields and feedstock properties. Additionally, the model integrates practical considerations from industrial operations, including multi-batch processing, virtual batch tracking, and customized inventory valuation, ensuring both its practicality and adaptability to real-world applications. Significant effort has been invested in data acquisition and processing, along with model formulation and validation. Three real-world case studies of different scales have been developed, with all model parameters fully accessible in \href{https://github.com/EMRPS/refinery-planning-benchmark}{\blue{\underline{EMRPS/refinery-planning-benchmark}}}.

Computational results indicate that existing off-the-shelf solvers are inadequate for effectively tackling industrial refinery planning problems. There remains a pressing need for the development of tighter relaxation models and more efficient techniques for generating cutting planes. Decomposition-based iterative algorithms also present a promising avenue for future research. Ablation experiments on delta-base modeling of key process units in an industrial case illustrate the significant impact of nonlinear relationships between product yields and feed properties on both the feasibility and optimality of solutions. Notably, this represents the first attempt to evaluate the feasibility of linearly approximated solutions established on actual processing routes. The proposed benchmark holds the potential to advance key areas within PSE, including algorithm development, decomposition mechanisms design, material value characterization, uncertainty management, and infeasibility diagnostics, offering a valuable resource for both academic research and industrial applications.


\section*{Data availability}
The sets, parameters, and computational results for the real-world cases can be found on GitHub (\href{https://github.com/EMRPS/refinery-planning-benchmark}{https://github.com/EMRPS/refinery-planning-benchmark}).

\nomenclature[A]{$T$}{set of periods}
\nomenclature[A]{$S$}{set of streams}
\nomenclature[A]{$S_P$}{set of products}
\nomenclature[A]{$S_M$}{set of raw materials}
\nomenclature[A]{$U$}{set of unit operations}
\nomenclature[A]{$U_{CDU}$}{set of CDUs}
\nomenclature[A]{$U_{PF}$}{set of process units without delta-base formulation}
\nomenclature[A]{$U_{PD}$}{set of process units with delta-base formulation}
\nomenclature[A]{$U_{MIX}$}{set of mixers}
\nomenclature[A]{$U_{SPL}$}{set of splitters}
\nomenclature[A]{$U_{BLD}$}{set of blenders}
\nomenclature[A]{$RB$}{set of proportional blending units}
\nomenclature[A]{$IU$}{set of pairs that stream $s$ is the input of unit $u$}
\nomenclature[A]{$OU$}{set of pairs that stream $s$ is the output of unit $u$}
\nomenclature[A]{$M$}{set of batches}
\nomenclature[A]{$IM$}{set of groups that stream $s$ is the input of unit $u$ in batch $m$}
\nomenclature[A]{$OM$}{set of groups that stream $s$ is the output of unit $u$ in batch $m$}
\nomenclature[A]{$SC$}{set of groups that cut fraction $s'$ can be merged into output stream $s$ of CDU $u$ in batch $m$}
\nomenclature[A]{$Q$}{set of properties}
\nomenclature[A]{$Q_V$}{set of volume-based properties}
\nomenclature[A]{$Q_W$}{set of weight-based properties}
\nomenclature[A]{$Q_P$}{set of percentage properties}
\nomenclature[A]{$CRU$}{set of properties under control for CDUs}
\nomenclature[A]{$SQ$}{set of traced property $q$ of stream $s$}
\nomenclature[A]{$FIX$}{set of fixed property $q$ of stream $s$}
\nomenclature[A]{$QT$}{set of property $q$ transferred from $s$ to $s'$}
\nomenclature[A]{$CDUMQ$}{set of property $q$ under control for batch $m$ of CDU $u$}
\nomenclature[A]{$DBSQ$}{delta-base formulation of unit $u$ in batch $m$ involves inlet stream $s$ and property $q$}
\nomenclature[A]{$VMQ$}{set of property $q$ under control for virtual batch $m$ of mixer-based unit $u$}
\nomenclature[A]{$C$}{set of capacity indices}
\nomenclature[A]{$CAPIN$}{set of inlet stream capacity control}
\nomenclature[A]{$CAPOUT$}{set of outlet stream capacity control}
\nomenclature[A]{$CAPS$}{set of stream $s$ under control for capacity $c$}
\nomenclature[A]{$USP$}{set of stream $s$ under composition control for process unit $u$}
\nomenclature[B]{$c^P_s$}{product price}
\nomenclature[B]{$c^M_s$}{raw material cost}
\nomenclature[B]{$ci^P_s$}{inventory product price}
\nomenclature[B]{$ci^M_s$}{inventory material cost}
\nomenclature[B]{$FQ^0_{s,q}$}{fixed property $q$ of stream $s$}
\nomenclature[B]{$FQ^{CRD}_{m,s,q}$}{property $q$ of crude oil $s$ in batch $m$}
\nomenclature[B]{$FQ^{CUT}_{m,s,s',q}$}{property transferring ratio for CDU}
\nomenclature[B]{$y_{u,m,s,s'}$}{cut fraction yield}
\nomenclature[B]{$\phi_{u,m,s,s'}$}{swing cut ratio}
\nomenclature[B]{$\gamma_{u,m,s}$}{base yield coefficient of stream $s$ in batch $m$ on unit $u$}
\nomenclature[B]{$B_{u,m,q}$}{baseline property $q$ of inlet stream $s$ on process unit $u$}
\nomenclature[B]{$\delta_{u,m,s,q}$}{unit step deviating from baseline yield coefficient of stream $s$ in batch $m$ of process unit $u$ for property $q$}
\nomenclature[B]{$\Delta_{u,m,q}$}{unit step deviating from baseline property $q$ in batch $m$ of process unit $u$}
\nomenclature[B]{$\alpha_{s,s',q}$}{property transferring ratio}
\nomenclature[B]{$w_{u,m,s,q}$}{component composition}
\nomenclature[B]{$\beta_{s,s'}$}{blending ratio}
\nomenclature[B]{$L^0_s$}{initial inventory level of stream $s$}
\nomenclature[B]{$[\underline{FV_{s,t}},\overline{FV_{s,t}}]$}{minimum/maximum mass flow of stream $s$ in period $t$}
\nomenclature[B]{$[\underline{FVC_{c,t}},\overline{FVC_{c,t}}]$}{minimum/maximum capacity in period $t$}
\nomenclature[B]{$[\underline{MFQ_q},\overline{MFQ_q}]$}{minimum/maximum property $q$ of the total crude oil processed in the system}
\nomenclature[B]{$[\underline{FQ_{s,q}},\overline{FQ_{s,q}}]$}{minimum/maximum property $q$ of stream $s$}
\nomenclature[B]{$[\underline{FQV_{u,m,q}},\overline{FQV_{u,m,q}}]$}{minimum/maximum property $q$ of batch $m$ in unit $u$}
\nomenclature[B]{$[\underline{FQB_{u,q}},\overline{FQB_{u,q}}]$}{minimum/maximum property $q$ for blender $u$}
\nomenclature[B]{$[\underline{L_{s,t}},\overline{L_{s,t}}]$}{minimum/maximum inventory level of stream $s$ during period $t$}
\nomenclature[B]{$[\underline{FC_{u,s}},\overline{FC_{u,s}}]$}{minimum/maximum feed composition proportion for stream $s$ of process unit $u$}
\nomenclature[C]{$X_{s,t}$}{equals 1 if the inventory level of stream $s$ decreases during period $t$}
\nomenclature[E]{$FVI_{s,t}$}{mass flow of inlet stream $s$ in period $t$}
\nomenclature[E]{$FVO_{s,t}$}{mass flow of outlet stream $s$ in period $t$}
\nomenclature[E]{$FVM_{u,m,s,t}$}{mass flow of stream $s$ in batch $m$ of unit $u$ in period $t$}
\nomenclature[E]{$FQ_{s,q,t}$}{property $q$ of stream $s$ in period $t$}
\nomenclature[E]{$\Gamma_{u,m,s,t}$}{calibrated yield coefficient of stream $s$ in batch $m$ of unit $u$ in period $t$}
\nomenclature[E]{$FVLI_{s,t}$}{mass flow of stream $s$ added to inventory in period $t$}
\nomenclature[E]{$FVLO_{s,t}$}{mass flow of stream $s$ taken from inventory in period $t$}
\nomenclature[E]{$L_{s,t}$}{inventory level of stream $s$ at the end of period $t$}
\nomenclature[E]{$VM_{u,m,s,t}$}{volume flow of stream $s$ in batch $m$ of unit $u$ in period $t$}
\nomenclature[E]{$V_{s,t}$}{volume flow of stream $s$ in period $t$}
\nomenclature[F]{CDU}{Crude distillation unit}
\nomenclature[F]{CRU}{Continuous reformer}
\nomenclature[F]{HRU}{Hydrogen recovery unit}
\nomenclature[F]{JHT}{Jet fuel hydrogenation unit}
\nomenclature[F]{HDC}{Hydrocracking unit}
\nomenclature[F]{DHT}{Diesel hydrogenation unit}
\nomenclature[F]{DCU}{Delayed coking unit}
\nomenclature[F]{HDT}{Hydrocracking and hydrotreating unit}
\nomenclature[F]{FCC}{Fluid catalytic cracking unit}
\nomenclature[F]{WHU}{Wax oil hydrotreating unit}
\nomenclature[F]{GHT}{Gasoline hydrogenation unit}
\nomenclature[F]{LDU}{Dry liquefied gas desulfurization unit}
\nomenclature[F]{GFU}{Gas fractionation unit}
\nomenclature[F]{RSU}{Reformate splitter}
\nomenclature[F]{ARU}{Aromatics complex}
\nomenclature[F]{C5S}{C5 separation unit}
\nomenclature[F]{ADU}{Aromatics disproportionation unit}
\nomenclature[F]{XFU}{Xylene fractionation unit}
\nomenclature[F]{DSU}{Dry gas separation unit}
\nomenclature[F]{ECF}{Ethylene cracking furnace}
\nomenclature[F]{PP}{Polypropylene unit}
\nomenclature[F]{CGH}{Cracked gasoline hydrogenation unit}
\nomenclature[F]{C9E}{C9 extraction unit}
\nomenclature[F]{EGU}{Ethylene glycol unit}
\nomenclature[F]{BEU}{Butadiene extraction unit}
\nomenclature[F]{MTBE}{MTBE unit}
\nomenclature[F]{PE}{Polyethylene unit}
\nomenclature[F]{EBU}{Ethylbenzene unit}
\nomenclature[F]{STU}{Styrene unit}
\nomenclature[F]{LP}{Linear programming}
\nomenclature[F]{MINLP}{Mixed-integer nonlinear programming}
\nomenclature[F]{MILP}{Mixed-integer programming}
\nomenclature[F]{NLP}{Nonlinear programming}
\nomenclature[F]{PSE}{Process systems engineering}
\nomenclature[F]{LPG}{Liquefied petroleum gas}
\nomenclature[F]{CGO}{Coker gas oil}
\nomenclature[F]{PE}{Polyethylene}
\nomenclature[F]{PP}{Polypropylene}
\nomenclature[F]{LGO}{Light gas oil}
\printnomenclature

\appendix


\bibliographystyle{elsarticle-num} 
\bibliography{cas-refs}





\end{document}